\documentclass[twocolumn]{aastex63}
\usepackage{xspace,color,soul,lineno}

\newcommand{\lya}{Ly\ensuremath{\alpha}\xspace}
\newcommand{\llya}{L(Ly\ensuremath{\alpha}\xspace)}

\shortauthors{Songaila et al.}

\begin{document}

\title{A Spectral Atlas of Lyman Alpha Emitters at $z=5.7$ and $z=6.6$}

\correspondingauthor{Antoinette Songaila}
\email{acowie@ifa.hawaii.edu}

\author[0000-0001-9028-6978]{A.~Songaila}
\affiliation{Institute for Astronomy, University of Hawaii,
2680 Woodlawn Drive, Honolulu, HI 96822, USA}

\author[0000-0002-6319-1575]{L.~L.~Cowie}
\affiliation{Institute for Astronomy, University of Hawaii,
2680 Woodlawn Drive, Honolulu, HI 96822, USA}

\author[0000-0002-3306-1606]{A.~J.~Barger}
\affiliation{Department of Astronomy, University of Wisconsin-Madison,
475 N. Charter Street, Madison, WI 53706, USA}
\affiliation{Department of Physics and Astronomy, University of Hawaii,
2505 Correa Road, Honolulu, HI 96822, USA}
\affiliation{Institute for Astronomy, University of Hawaii, 2680 Woodlawn Drive,
Honolulu, HI 96822, USA}

\author[0009-0008-7427-4617]{E.~M.~Hu}
\affiliation{Institute for Astronomy, University of Hawaii,
2680 Woodlawn Drive, Honolulu, HI 96822, USA}

\author[0000-0003-1282-7454]{A.~J.~Taylor}
\affiliation{Department of Astronomy, The University of Texas at Austin,
2515 Speedway, Austin, TX 78712, USA}
\affiliation{Department of Astronomy, University of Wisconsin-Madison,
475 N. Charter Street, Madison, WI 53706, USA}



\begin{abstract}
We present two uniformly observed spectroscopic samples of \lya\ emitters (LAEs)
(127 at $z=5.7$ and 82 at $z=6.6$), which we use to investigate the evolution of the LAE population
at these redshifts.
The observations cover a large field (44~deg$^2$) in the North Ecliptic Pole (HEROES),
as well as several smaller fields.
We have a small number of exotic LAEs in the samples: double-peaked \lya\
profiles; very extended red wings; 
and one impressive lensed LAE cross. We also find three broad-line AGNs.
We compare the \lya\ line width measurements at the two redshifts, finding that the
lower-luminosity LAEs show a strong evolution of decreasing line
width with increasing redshift, while the high-luminosity LAEs do not, with a
transition luminosity of $\log$~L(\lya)~$\approx43.25$~erg~s$^{-1}$.
Thus, at $z = 6.6$, the high-luminosity LAEs may be producing large
ionized bubbles themselves, or they may be residing in overdense galaxy sites that
are producing such bubbles. In order to avoid losses in the red wing,
the radius of the ionized bubble must be larger than
1~pMpc. The double-peaked LAEs also require transmission on the blue side.
For the four at $z=6.6$,
we use models to estimate the proximity radii, $R_{a}$, where the ionizing flux
of the galaxy is sufficient to make the surroundings
have a low enough neutral fraction to pass the blue light.
Since the required $R_a$ are large, multiple ionizing sources
in the vicinity may be needed.
\end{abstract}



\section{Introduction}

The epoch of reionization is a key time in the evolution of the Universe,
when the intergalactic medium (IGM) transitioned from being neutral to being very highly ionized.
As we discuss below, it is likely that reionization is patchy and that
the ionization begins with  HII regions around the ionizing sources, which then expand and
merge. However, our understanding is still quite limited. Intensive 
simulations
are currently being carried out to model this evolution, but there are many issues
that can only be treated in an empirical fashion. Perhaps most critically, we still do
not know which objects were the source of the photons that reionized the IGM, or the exact redshift
range when reionization occurred.
Answering these questions are major goals in observational cosmology and
fundamental for simulating galaxy evolution.

\begin{figure*}[ht]
\centerline{
\includegraphics[width=9.5cm]{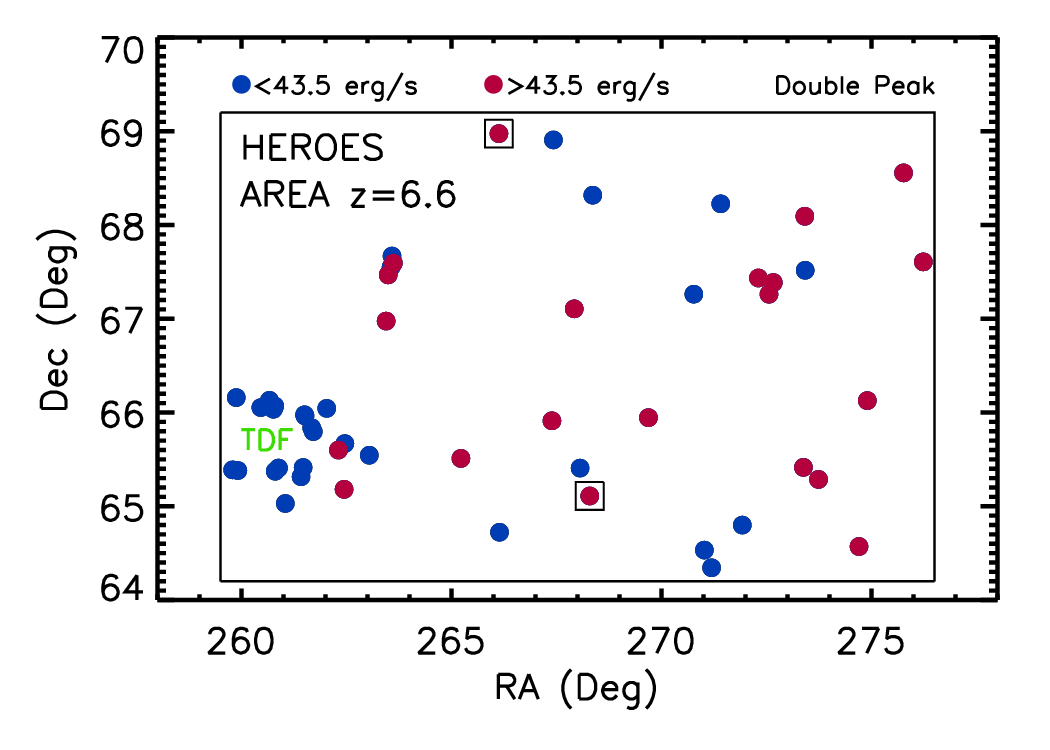}
\includegraphics[width=8.5cm]{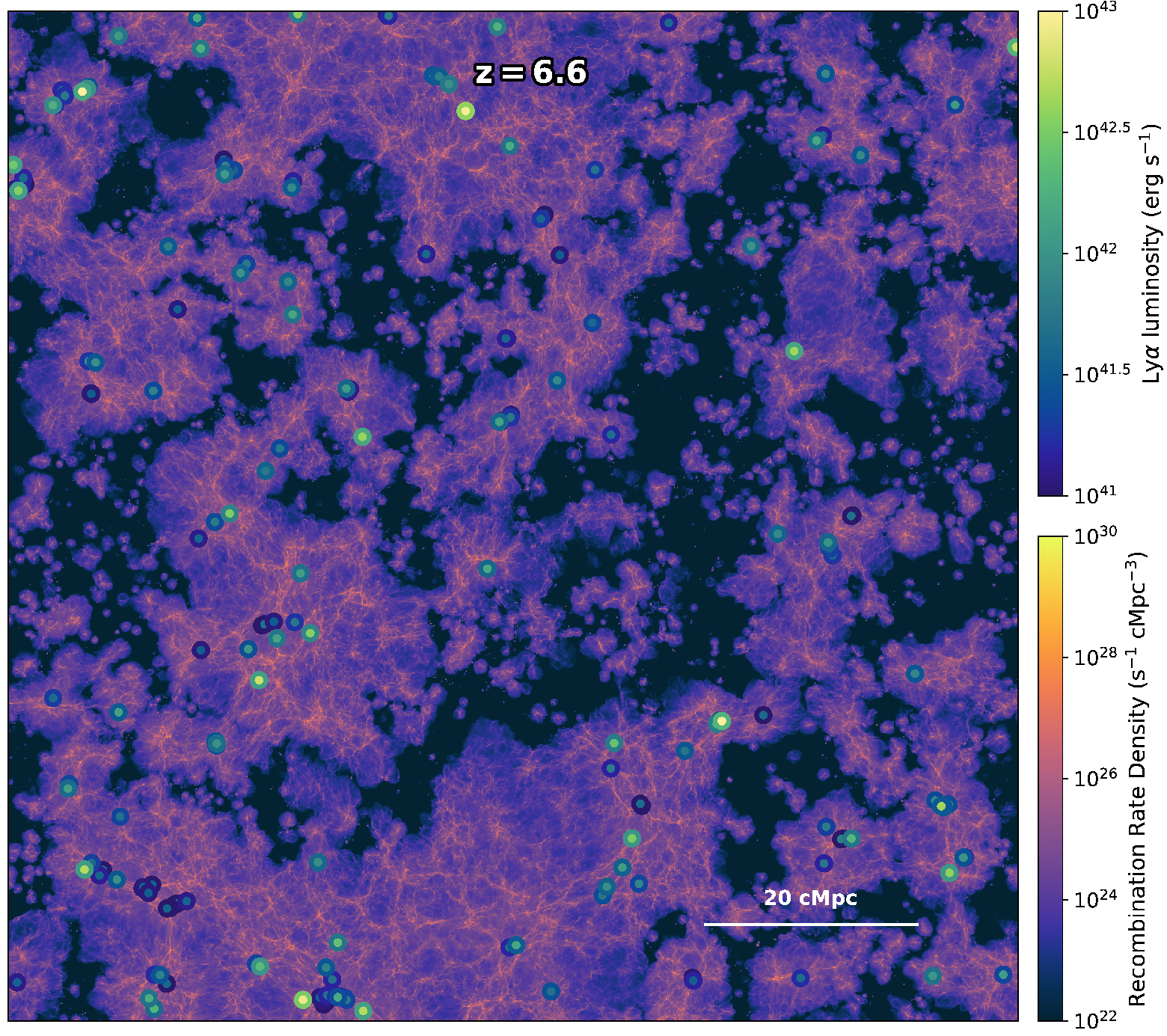}
}
\caption{
(Left) Positions of spectroscopically identified LAEs at  $z=6.6$ 
in the HEROES field (data from \citealt{taylor23} and work in progress).
The red points show the 21 sources with $\log$ L(\lya) $>43.5$~erg~s$^{-1}$, where
the sample is complete.
The two sources with double-peaked spectra are marked with open squares.
The sources in the lower left lie in the area around the JWST Time Domain Field
(\citealt{jansen18}), where our deeper HSC NB921 observations allow us to
probe the lower luminosity $z=6.6$ LAE population (labeled TDF in green).
(Right) Simulations such as THESAN (figure from \citealt{xu23}; circles show
LAEs at $z=6.6$, and purple shading shows the ionization of the IGM) 
are now reaching the point where we
can make direct comparisons with observations of the numbers and spatial distribution
of LAEs. The portion of the THESAN simulation shown
is almost exactly matched in area 
and redshift coverage to HEROES.
\label{radec}
}
\end{figure*}

There has been enormous recent progress in identifying populations of high-redshift galaxies, 
particularly \lya\ emitters (LAEs) at $z\gtrsim6$. These samples have the potential
to constrain when reionization occurred, since both
the strength and shape of the \lya\ line can be modified by the radiative damping wings of neutral
 hydrogen in the IGM 
(e.g., \citealt{haiman02,haiman05,mason20}). As the neutral
hydrogen fraction in the IGM increases with increasing redshift, we
expect the \lya\ emission lines of LAEs to become narrower and less luminous, and only the red wings of the lines
to be visible. At the highest redshifts,
we will no longer see the \lya\ line at all, except in the most highly ionized regions of the IGM.

This evolution of the LAE population is apparent in both a drop in the normalization of the LAE 
luminosity function (LF)
towards higher redshifts and a decrease in the LAE line widths
from $z=5$ to $z=7$, though both of these results appear to depend on the luminosities
of the LAEs (\citealt{konno14,konno18,santos16,itoh18,
taylor20,taylor21,goto21,ning22,songaila22}).

A similar result is seen in UV continuum samples, where the fraction of objects
with strong Ly$\alpha$\ emission appears to decline beyond $z=6$
(\citealt{stark10,pentericci18,hoag19}). However,  both the significance of this result
and the redshift where
the drop occurs are controversial (see, e.g., \citealt{stark16,fuller20,kusakabe20}).

Quasar spectra support the idea of patchiness in the reionization process
(e.g., \citealt{becker15}), as do other lines of
evidence, such as multi-peaked LAE spectra (\citealt{hu16,songaila18,bosman20,meyer21}) and
the distribution of LAEs.
As an  example of the latter effect, \citet{larson22} found a candidate LAE at $z = 8.7$
near a known source at a similar redshift (\citealt{zitrin15}).
Furthermore, we are beginning to see
LAEs well into the redshift range  where we would expect the IGM to
be substantially neutral (e.g., \citealt{oesch15, zitrin15,
hoag18, hashimoto18, pentericci18}). Indeed, with the advent of JWST,  
Ly$\alpha$\ emission has now been detected out to
$z=10.60$ (GN-z11; \citealt{bunker23}). 

In combination, all of these observations can be matched to models of the evolution
of LAEs and used to constrain the evolution of the IGM.
Extensive simulations combining sophisticated radiative transfer codes with cosmological
models, such as THESAN (e.g., \citealt{garaldi22,neyer23}) and SPHINX (e.g., \citealt{garel21,
katz23}), are now being carried out for this purpose. Despite the beautiful results from these 
simulations, there are still many uncertainties.
Perhaps most importantly,  it is not easy  to disentangle the effects of the IGM  from the complexity
 of the galaxy's Ly$\alpha$\ emission and the scattering at circumgalactic scales
(\citealt{laursen11,jensen13,garel15,hassan21,guo23}).

In order to model the evolution of the IGM at very high redshifts,
we need to separate  the effects of the IGM from the intrinsic structure
of the Ly$\alpha$ line emerging from the galaxy and its surrounding
circumgalactic medium (CGM). The simplest approach is to assume
that the emergent Ly$\alpha$ lines are varying more slowly 
with redshift than the IGM properties.
However, even with such a crude assumption, 
we must have a large reference sample to
understand how the significant variations in 
the emergent Ly$\alpha$ lines (e.g., \citealt{guo23})
combine with the IGM structure to match to the observations. More
sophisticated modeling also requires detailed information on large
samples of LAEs to constrain the underlying assumptions.
A further complication is that the line properties appear to  be a function
of the \lya\ luminosity (e.g., \citealt{santos16,taylor20,taylor21,ning22,songaila22}),
introducing a shape dependence in the evolution of the \lya\ LF
and a dependence of the velocity width on the \lya\ luminosity.
Thus, we also need to sample a wide range of \lya\ luminosities.

In practice, the highest redshift where high-quality spectra of hundreds of LAEs
can be straightforwardly obtained is $z=7$, where giant optical imagers, 
such as Hyper Suprime-Cam (HSC) on the Subaru 8.2~m telescope, 
can be used to generate narrowband selected samples
(e.g., SILVERRUSH: \citealt{konno18}; HEROES: \citealt{taylor23};
see also the LAGER survey, which uses DECam on the CTIO Blanco
4~m telescope: \citealt{wold22}).
These can then be followed up
with efficient optical spectrographs to generate large spectroscopic samples.
The lowest background long-wavelength
regions in the atmospheric emission are at 8160~\AA\ and 9210~\AA,
which correspond to $z=5.7$ and $z=6.6$;
$\sim 100$~\AA\ filters at these wavelengths are used 
to make the initial selections of narrowband excess objects
(see, e.g., \citealt{shibuya18,ning22,kikuta23,taylor23}).
Ideally, the LAE samples should be complete to a 
limiting observed \lya\ luminosity and cover large contiguous areas
(10s of deg$^2$) in fields with extensive ancillary observations.

Our primary field is the 44~deg$^2$  HEROES field in the North
Ecliptic Pole (NEP; \citealt{taylor23}), which has, or will have, deep coverage
from the eROSITA X-ray and SPHEREx spectroscopy missions and
contains the largest of the Euclid Deep Fields (20~deg$^2$).
The goal of the present paper is to provide a publicly available atlas of uniform spectra 
for a very large sample of LAEs at $z=5.7$ and $z=6.6$ over a range of observed
\lya\ luminosities in this field together with a number of smaller
areas. This sample can be used to characterize the LAE properties,
to assess the evolution of the ionization structure, and to match to the models 
at these later stages in the reionization process. 
As an example, in Figure~\ref{radec}, we show the current spectroscopically
confirmed LAE sample at $z = 6.6$ in HEROES, which we
compare with a similarly sized  area in the THESAN simulation
of \lya\ emission at high redshifts (\citealt{xu23}). In this simulation,
the $z=6.6$ LAEs are seen to occur in regions of the IGM that are more ionized (purple shading)
and to be absent in regions that are more opaque.

We present our spectroscopic observations and discuss our observed
\lya\ line profiles in Section~\ref{obs}. In Section~\ref{exotic},  we briefly
summarize the various types of exotic LAEs seen in the sample.
In Section~\ref{widthmeasurements}, we make our \lya\ line width measurements.
In Section~\ref{vel_wid}, we discuss the evolution
of the velocity widths with redshift. 
Finally, in Section~\ref{discuss}, we consider
the constraints on the IGM evolution that may be derived from the observations.

We use a  $H_0=70$~km~s$^{-1}$~Mpc$^{-1}$, $\Omega_M=0.3$, 
and $\Omega_{\Lambda}=0.7$ cosmology throughout.

\section{Observations}
\label{obs}

\begin{figure}
\centerline{
\includegraphics[width=9.5cm]{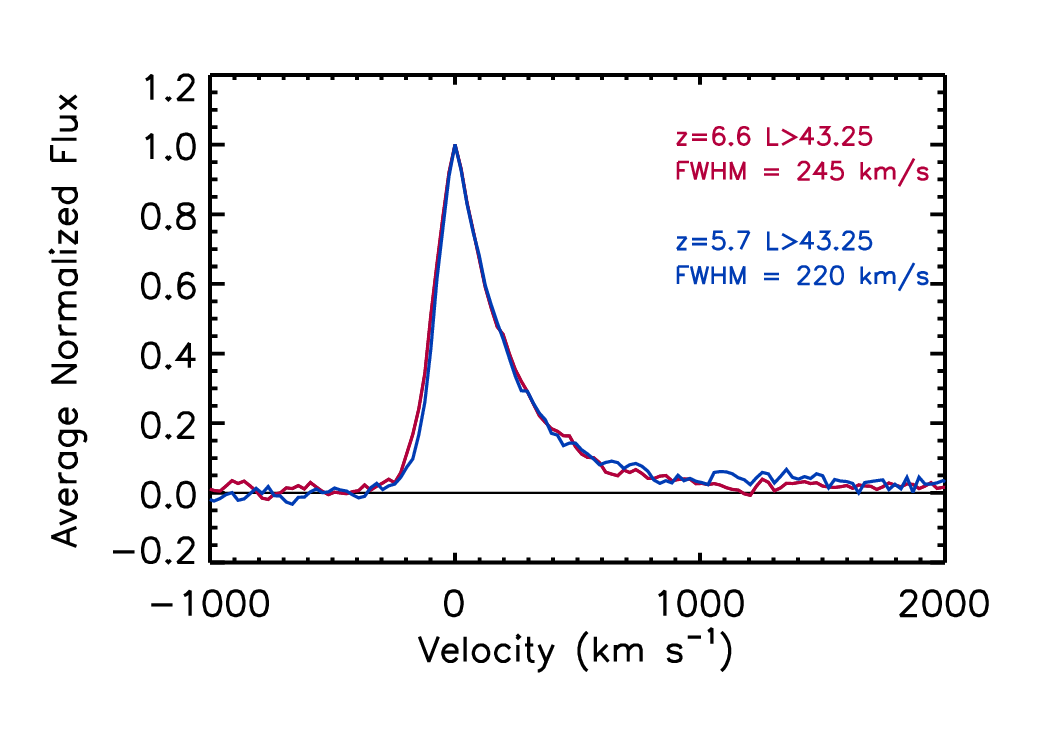}
}
\vskip -0.85cm
\caption{Average normalized  \lya\ spectra 
for luminous LAEs ($\log$ L(\lya)$>43.25$ erg~s$^{-1}$)
at $z=5.7$ (blue; 55 objects) and
$z=6.6$ (red; 57 objects). The FWHM of the
stacked spectra at the two redshifts are shown. 
The spectra are from the 830 line~mm~$^{-1}$ grating on the 
Keck/DEIMOS spectrograph. The instrumental FWHM
of 85~km~s$^{-1}$ at $z=6.6$ and 96~km~s$^{-1}$ at $z=5.7$ are not subtracted here.
\label{stacks}
}
\end{figure}

We have been obtaining follow-up spectroscopy of narrowband selected samples 
(using the narrowband filters NB816 for $z=5.7$ and NB921 for $z=6.6$) in HEROES
and other fields in the Subaru/HSC and Subaru/Suprime-Cam archives.
These Subaru imaging observations provide very large samples (many thousands) of 
photometrically selected narrowband
excess sources satisfying the LAE selection criteria:  NB816$-i >1.2$
for $z=5.7$, with the sources also undetected in the $g$ and $r$ bands,
and NB921$-z >1$ for $z=6.6$, with the sources also undetected in the
$g$, $r$, and $i$ bands.
Thus, we have many available candidates for follow-up spectroscopy. 
Most of these lie in small areas (5--7~deg$^2$). HEROES (Figure~\ref{radec})
is the largest contiguous field (44~deg$^2$) in our spectroscopic survey.

We give the observed field for each object in Tables~1, 2, and 3, along with the
R.A. and Decl. of each object. The selection of the sources
observed prior to 2010 is summarized in \citet{hu10}. 
For sources observed after 2010, we give the narrowband and continuum magnitudes 
in Table~3, together with cutouts of the images when these were observed with HSC. 
The LAEs  in the NEP (52 of the $z=6.6$ sample and 34 of
the $z=5.7$ sample) were taken  from the catalog of \citet{taylor23}.
The remaining sample comes from the SSA22, XMM-LLS, COSMOS, and
GOODS-N fields, all of which were observed as part of the HSC-SSP initiative \citep{aihara22}.

\begin{figure*}[th]
\centering
{\includegraphics[width=1.5\columnwidth]{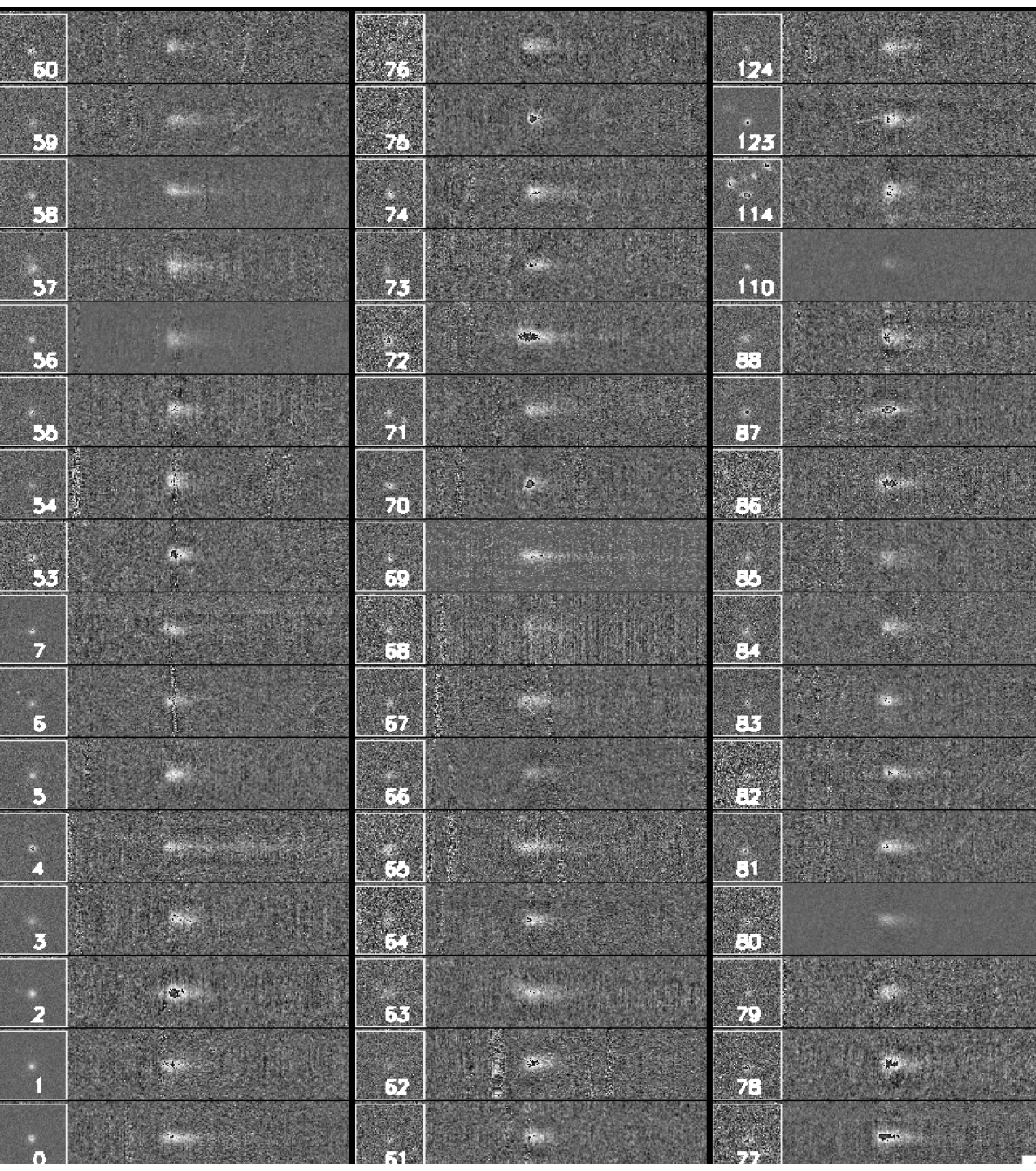}
}
\vskip -1cm
\caption{A subset of our existing
LAE spectra at $z=5.7$. We show a narrowband image thumbnail for each LAE (numbered),
along with its 2D spectrum, where the $x$-axis corresponds to the wavelength,
and the $y$-axis corresponds to the spatial position along the slit. The $x$-axis
is 140~\AA\ long, and the $y$-axis is $7\farcs2$.
}
\label{contspec}
\end{figure*}

\begin{figure*}
\centerline{\includegraphics[width=9.25cm,angle=0]{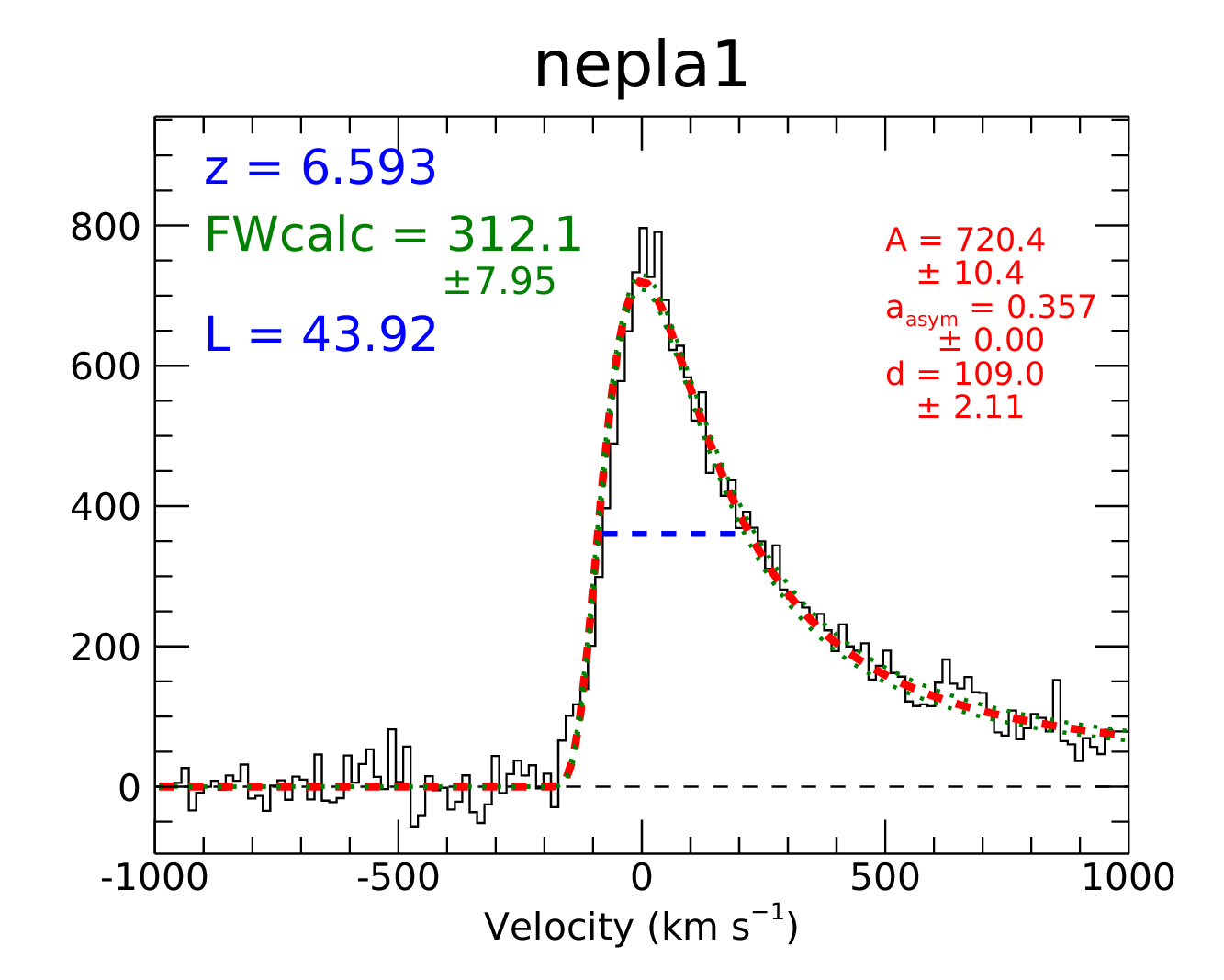}
\includegraphics[width=9.25cm,angle=0]{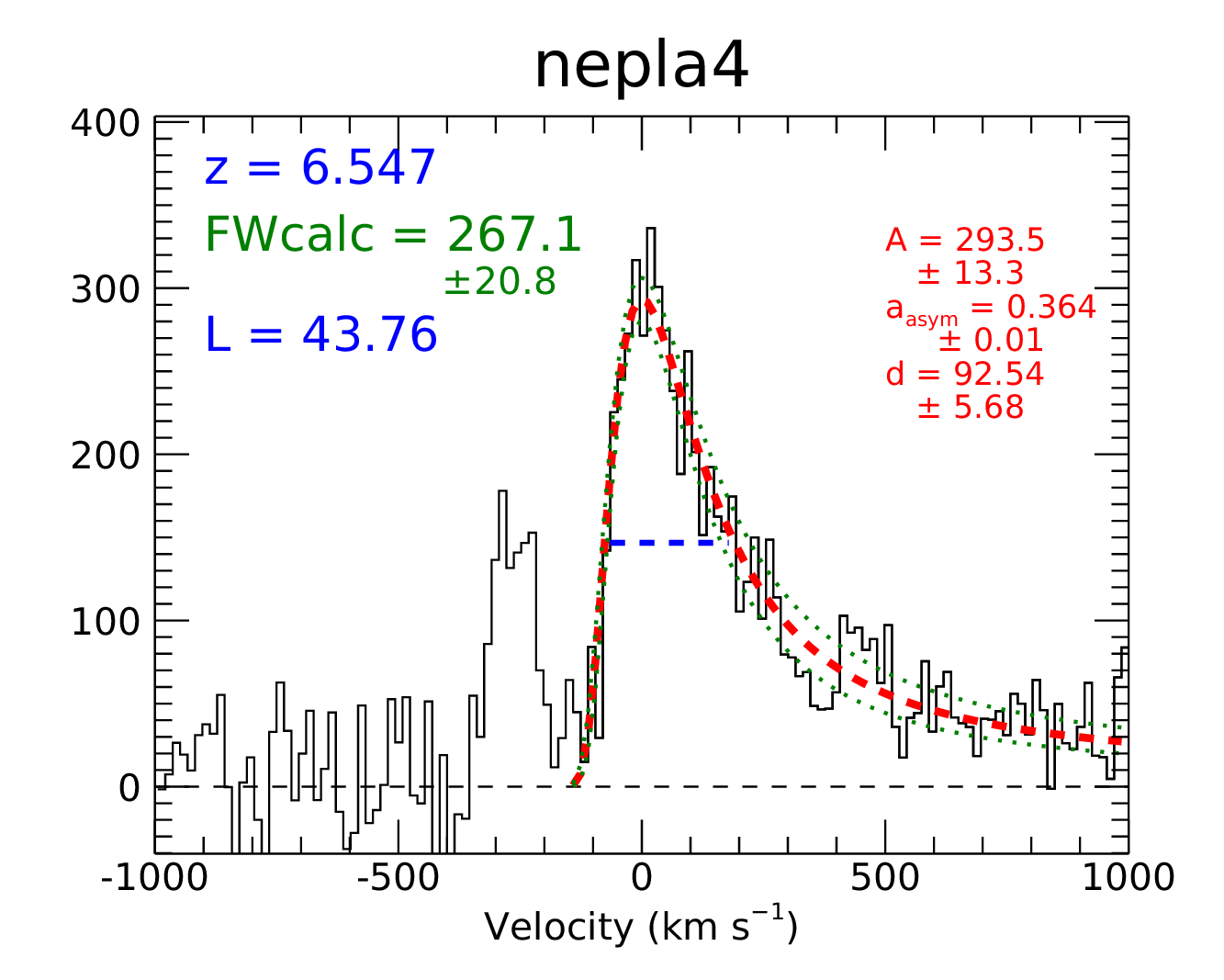}}
\caption{Asymmetric fits from Section~\ref{widthmeasurements} for two sources
in HEROES (left: NEPLA1, a single-peaked LAE; right: NEPLA4, 
a double-peaked LAE, where the
second peak is blueward of the principal \lya\ peak).
The red dashed line shows the fit to the data (black).
The FWHM in km~s$^{-1}$ based on the fit is given in green
and shown as the blue dashed line. The redshift and
$\log$~L(\lya) are given in blue, and the fitting parameters from
Equation~\ref{eq_asymgauss} in red.  The $y$-axis shows the flux
in arbitrary units. 
}
\label{nb921_skew}
\end{figure*}

We carried out our spectroscopic observations over a very extended period
from 2002 to August 2023, but the observations were all done in a uniform fashion
with the same spectrograph configuration and wavelength range.
We give the date of each observation in Table~3.
\citet{hu10} described some of the earlier spectra,
and spectroscopic observations for a number of the sources
have appeared in the literature (e.g., \citealt{sobral15,jiang17,matthee17,shibuya18,songaila18,songaila22}).

We used  the Keck/DEIMOS spectrograph with the 830~line~mm$^{-1}$ grating
and $1''$ slits, which gives good red sensitivity and moderately high resolution.
Because of ghosting in this configuration, we used three dithered 20~min observations.
The total exposure time is 1~hr for most of the sources.
Some LAEs have multiple observations and hence longer exposure times. 
The total exposure time for each source may be found in Table~3. 
Because the sources are sparse, there are usually only one or two LAEs in each DEIMOS mask.
The data reduction is described in \citet{cowie96}. We measure the instrumental
resolution from the sky lines, giving an average of 85~km~s$^{-1}$ at $z=6.6$ and 
96~km~s$^{-1}$ at $z=5.7$. 

We summarize the LAE targets and the fields they are drawn from
in Tables~\ref{z5tab} ($z=5.7$) and \ref{z6tab} ($z=6.6$).
We give the signal-to-noise of the LAEs in Table~3.

On average, the LAEs at these high redshifts have a very generic (and well-known) spectral
shape, with a sharp break at the blue side and a tail to the red side (see Figure~\ref{stacks}).
The small number of low-redshift contaminants of the narrowband candidate sources ($\sim5$\%) 
are easily recognized by the
doublet nature of the [OII]3727 and [OIII]5007 lines and were dropped from the
sample. A larger number of the narrowband selected sources ($\sim25$\%)
were not confirmed by the spectroscopy and were also eliminated.
The weak continuum also shows the break at the \lya\ wavelength. 
We see very little change in the shape and width of the average \lya\ line profile
between $z=5.7$ and $z=6.6$.
The shape is governed by the escape of \lya\ from the galaxy and its circumgalactic gas,
combined with the blue-side scattering from the neutral IGM. (See \citealt{guo23} and references therein 
for an extensive recent discussion.)

However, these average line profiles are, in reality, drawn from 
a fairly diverse set of individual line shapes and widths, as we
illustrate in Figure~\ref{contspec}, where we show two-dimensional (2D) spectra
for a subset of our $z=5.7$ LAEs. We see objects with a wide range
of red velocity wings, including some spectacularly long-tailed
objects, such as sources~4 and 69.
We also see objects with blue emission relative to the \lya\ peak,
such as source~2, and lensed systems, such as source~114.
This variety reflects the variation in the redshift of the galaxy
with respect to the local IGM, the variation in the ionization surrounding
the source, and the degree to which the escape from the galaxy and the CGM
has moved the emission to the red.

\begin{figure*}
\centerline{
\includegraphics[width=0.5\linewidth]{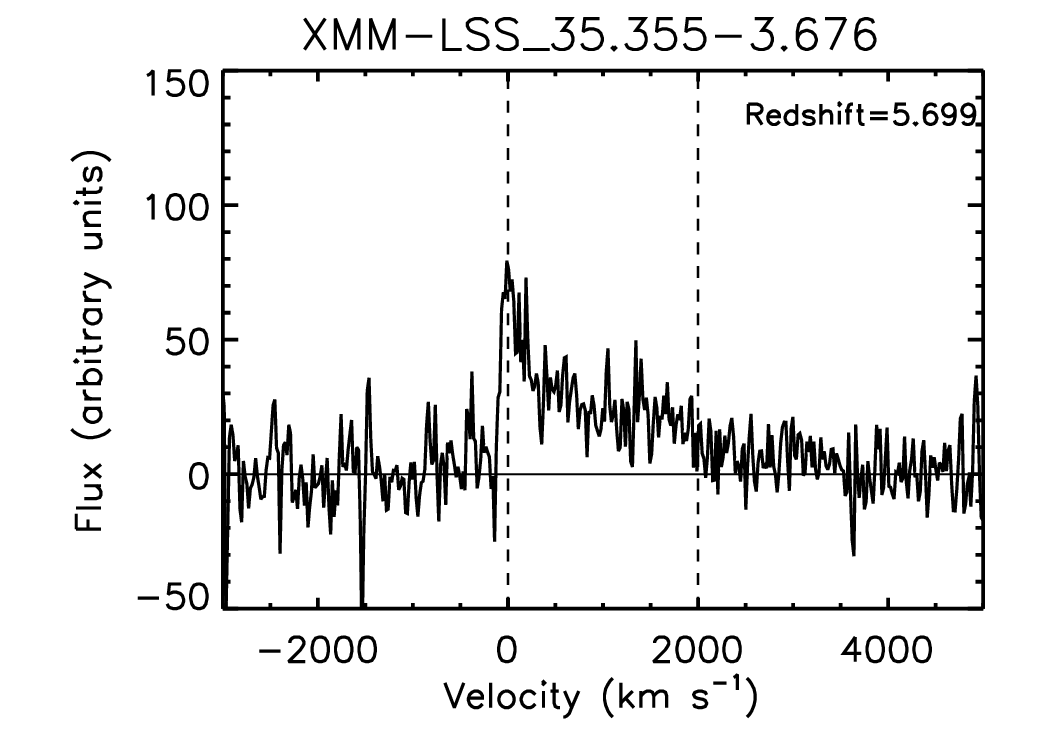}
\includegraphics[width=0.5\linewidth]{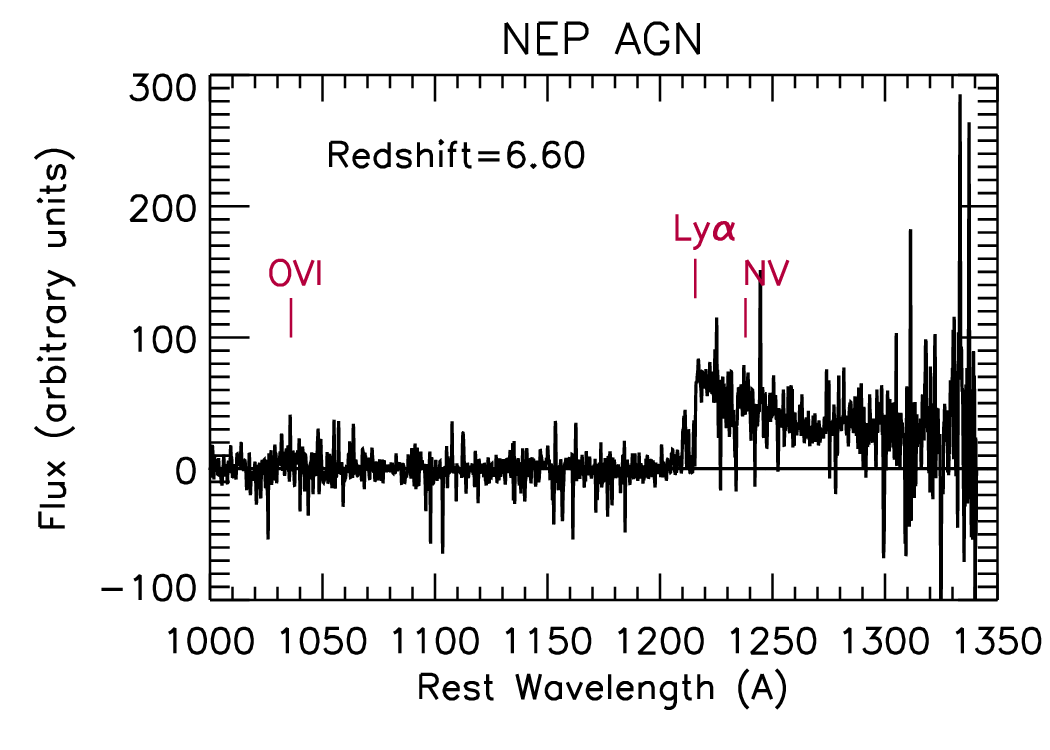}
}
\vskip -0.5cm
\caption{
(Left) An LAE with an extended red wing.
In addition to a narrow line, the red wing extends
to $\sim3000$~km~s$^{-1}$. The dashed lines
show zero velocity and $2000$~km~s$^{-1}$.
(Right) A broad-line AGN detected in the \lya\ selected sample.
}
\label{longwing}
\end{figure*}

\begin{figure}[ht]
\begin{centering}
\includegraphics[width=1.0\linewidth]{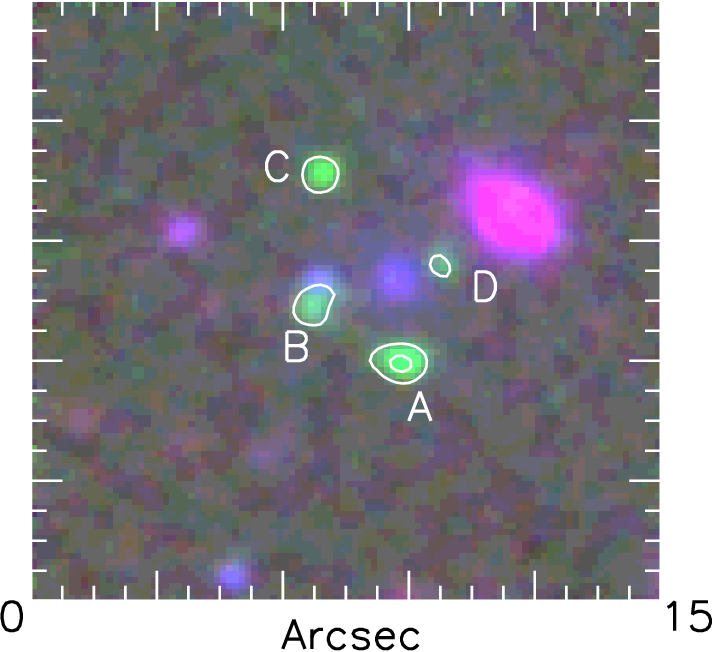}
\vskip 0.2cm
\includegraphics[width=0.90\linewidth]{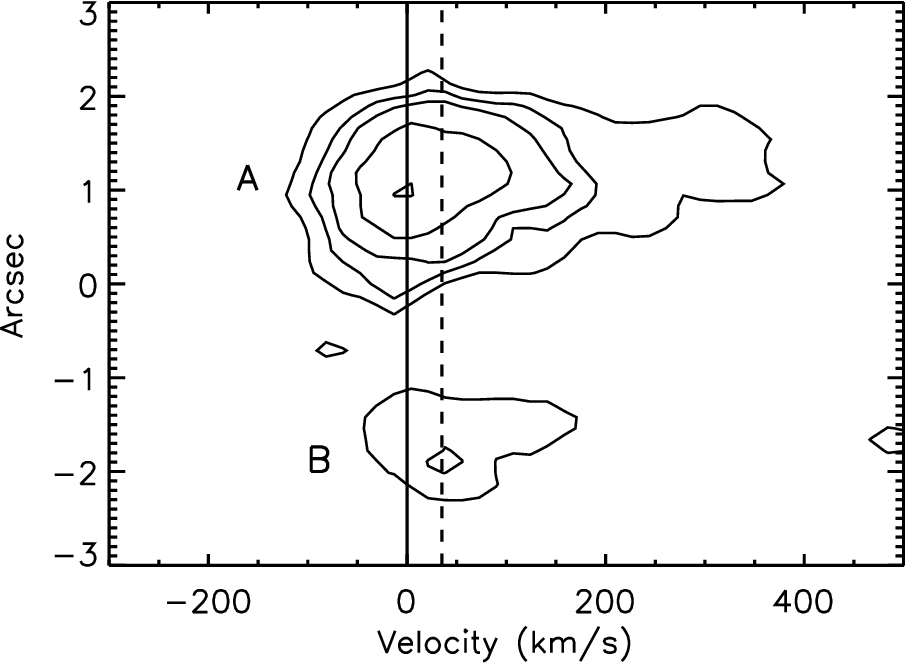}
\caption{
Upper panel: Structure of a $z=5.749$ lensed LAE in the SSA22 field.
Green indicates \lya\ emission
based on NB816 imaging from Subaru/HSC. 
For the remaining galaxies, blue shows
$r$-band and red shows $z$-band. The
differential paths allow us to investigate
the small-scale structure in the CGM and IGM.
Lower Panel: 2D spectra from 
a slit covering components B and A. The
$x$-axis corresponds to the velocity relative
to the peak of component A. The solid line shows
zero velocity, and the dashed line shows 35 km~s$^{-1}$, which corresponds to the
peak of component B.
}
\label{fig:la_cross}
\end{centering}
\end{figure}

\section{Exotic LAEs}
\label{exotic}
As mentioned above, our large LAE spectroscopic samples allow us to
find rare and interesting objects. In general,
only one to a few of these have been found, 
emphasizing the need to continue to increase
the numbers in our spectroscopically observed sample
so we can find further examples and make statistical use of them.
We describe three categories of such exotic LAEs here.

\subsection{Double-peaked LAEs}
While double-peaked \lya\ profiles resulting from the escape of the line 
from the galaxy and the CGM are seen in $\sim$30\% of low 
to moderate redshift LAEs (e.g., \citealt{kulas12}), at high redshifts,
we expect any blue wing to be removed by scattering in the IGM. 
However, we see a handful of LAEs with blue-side emission
(e.g., \citealt{hu16,songaila18,bosman20,meyer21} for published examples).
We show a double-peaked LAE in Figure~\ref{nb921_skew}, 
which we compare with a single-peaked LAE of similar luminosity.

Double-peaked \lya\ emission may be direct evidence
for large ionized bubbles around these high-redshift LAEs, which may reveal reionization in progress. 
In this interpretation, we are seeing the \lya\ profile emerging from the galaxy directly without scattering by the IGM.
However, we must also consider the possibility that double-peaked \lya\ profiles could result from processes
other than ionized bubbles. \citet{matthee18} suggested that they could be the result of a merger, where 
both galaxies are producing \lya\ lines redward of the IGM cutoff but with different velocities. 
This could then result in an observed double-peaked structure with different components in the 
LAE having significantly different redshifts.

\subsection{Extended red wing LAEs and AGNs}
As we noted in Section~\ref{obs},
we see a small number of LAEs with very extended
red wings, in some cases to velocities beyond 2000 km s$^{-1}$.
\citet{yang14} previously discussed extended red wings, but the
object they considered only extends to 1000~km~s$^{-1}$. The present
sample contains much more extreme objects. The $z = 5.7$ sample
contains 3 sources (4, 45, and 69) with tails extending beyond
1500~km~s$^{-1}$. At $z=6.6$,
there are two such sources (35 and 43). Source 35 is also the double-peaked source
NEPLA4. 

There are also two active galactic
nuclei (AGNs) in the $z=6.6$ redshift range (Figure~\ref{longwing}, right panel; see also
Figure~\ref{z6_groups}) and one in the $z=5.7$ range.
These sources are not included in the LAE samples.

In Figure~\ref{longwing} (left panel),
we show the spectrum of one of the extended red wing LAEs (source 4 in the $z=5.7$ sample). 
In addition to a narrow component, the spectrum shows a red wing extending to
beyond 3000~km~s$^{-1}$. It is difficult to reproduce such extreme line widths with models 
of the Ly$\alpha$ escape from star-forming galaxies. One possible alternative interpretation
is that we are seeing a combination of narrow \lya\ emission
from the galaxy and broad \lya\ emission from an AGN in the galaxy. 
These sources might be related to JWST-detected AGNs at these redshifts, some of which  
also show narrow lines with underlying broad components,
but in the rest-frame optical (\citealt{kocevski23,harikane23,greene24}).

\subsection{Lensed LAEs}
Lensed LAEs are of particular interest, since the differential
paths allow us to probe for the presence of small-scale
structure in the circumgalactic and intergalactic gas.
So far, we have only found one such object: an LAE cross
at $z=5.749$ in the SSA22 field (see Figure~\ref{fig:la_cross}, upper panel). 
The components of the cross lie at separations from $2\farcs5$ to $5\farcs2$, 
and while the components with measured redshifts (A and B)
lie at similar redshifts, their \lya\ profiles are not identical.
Thus,  as we show in Figure~\ref{fig:la_cross} (lower panel),
component B is slightly (but significantly)
redward of component A by $\sim35$~km~s$^{-1}$.
These changes in the \lya\ profiles may reflect changes in
the \lya\ scattering for scales of order 10~parsec in the
CGM to several hundred parsec in the IGM.
The variations in the \lya\ profiles should provide
constraints on the modeling of LAEs and on how the profiles
change with viewing angle (\citealt{guo23}).

We will consider the double-peaked sources further
in the discussion, but we postpone consideration of the
extended red wing sources and the LAE cross to subsequent papers.

\section{\lya\ Line Width Measurements}
\label{widthmeasurements}
For each LAE, we first determine the redshift corresponding to
the peak of the \lya\ line. We then make a parameterized fit to determine the
FWHM in velocity.
This gives an accessible measurement of the
properties of the LAE and allows for a simple classification.

Following \citet{shibuya14} and \citet{claeyssens19},
we fit the \lya\ line with an asymmetric Gaussian profile,
\begin{equation}\label{eq_asymgauss}
        f(\Delta v) = A \exp \left(-\frac{\Delta v^2}{2(a_{\rm asym} (\Delta v) +d)^2}\right) \,,
\end{equation}
where $A$ is the amplitude (normalization),
and $\Delta v$\ is the velocity relative to the peak of the 
\lya\ profile.  Here $a_{\rm asym}$ controls the asymmetry, and $d$ controls the line width.
In terms of these free parameters, the width of the line (in km~s$^{-1}$) is
\begin{equation}\label{eq_asymwidth}
       {\rm FWHM}= \frac{2 \sqrt{2 \ln 2}\  d}{(1-2\ln 2 \ a_{\rm asym}^2)} \,.
\end{equation}
For the double-peaked sources, we fit only to the red wing of the LAE
(right panel of Figure~\ref{nb921_skew}).

As can be seen from the two examples shown in Figure~\ref{nb921_skew},
the asymmetric fit generally 
provides an extremely good representation of the line.
At both redshifts, all of the asymmetry parameters are positive,
corresponding to the wings being red. The median asymmetry parameter
is 0.26 at $z=5.7$ and 0.28 at $z=6.66$, and the denominator in Equation~2 increases the FWHM
by $\sim 10$\%, on average. The corresponding error in the
FWHM is nearly fully dominated by the error in $d$, with only
a few percent contribution from the $a_{asym}$ term. 
We tabulate the redshifts, \lya\ luminosities, FWHM, and asymmetry parameters in 
Tables~\ref{z5tab} and \ref{z6tab}.

\begin{figure}
\centerline{\includegraphics[width=10cm]{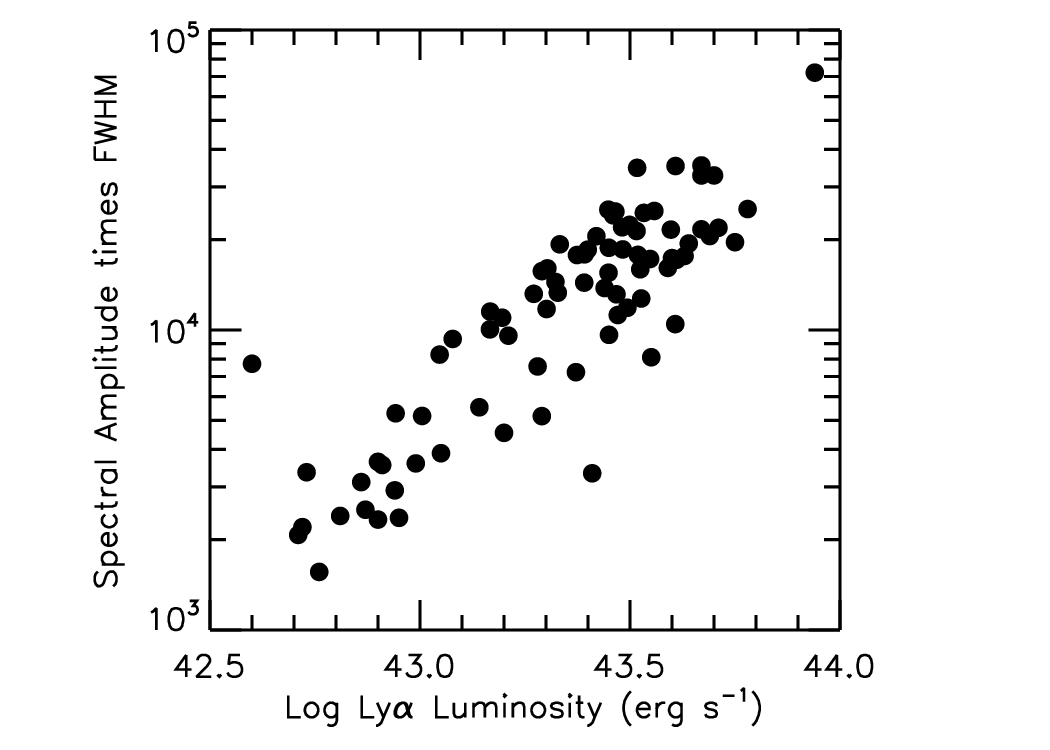}
}
\caption{Comparison of the spectral line strength as
measured by $A\times$FWHM vs. \llya\  for the $z=6.6$ LAEs. 
Only sources with measured signal-to-noise $>5$ in the amplitude are shown.
}
\label{ampl_test}
\end{figure}

\begin{figure}
\centerline
{\includegraphics[width=9.25cm]{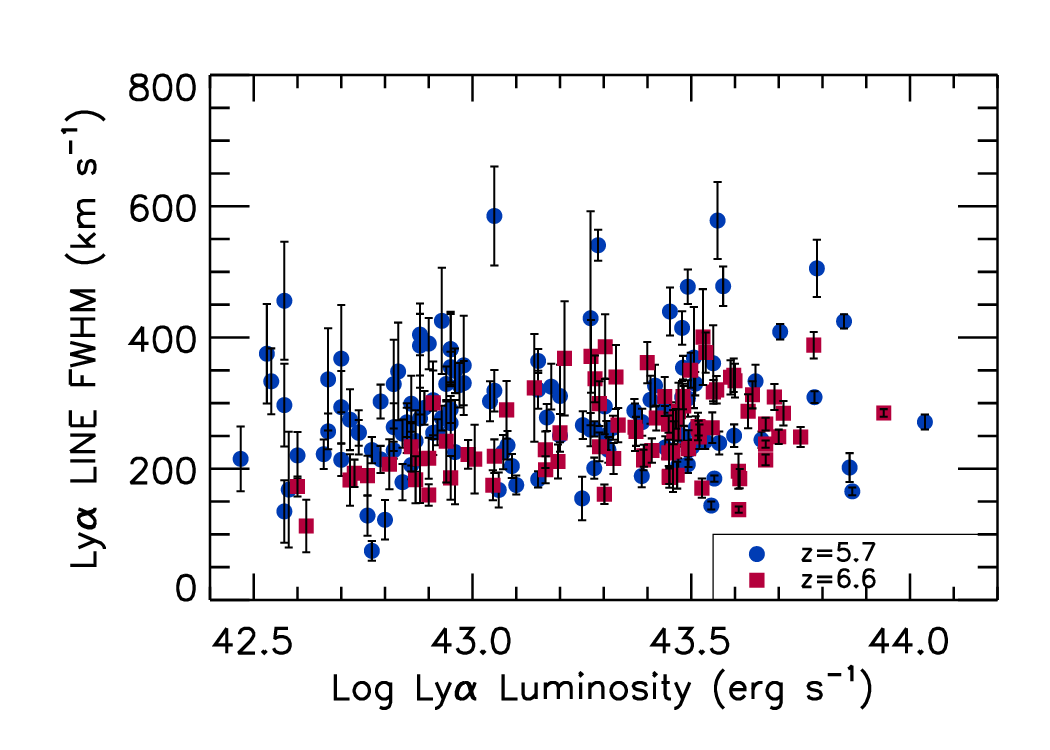}
}
\caption{
\lya\ line width vs. \llya\ for our $z = 5.7$ (blue circles) and $z = 6.6$ (red
squares) LAEs. The lower-luminosity LAEs 
show a strong evolution of decreasing line
width with increasing redshift, while the high-luminosity LAEs do not. The
transition luminosity lies at $\log$~L(\lya)~$\approx43.25$~erg~s$^{-1}$.
The errors are 68\% confidence on the FWHM.
}
\label{LF_LW}
\end{figure}

We next calculate the \lya\ line fluxes and luminosities. Because
of the difficulty of an absolute calibration for slit spectra, we calculate the 
\lya\ line fluxes from the narrowband imaging. 
We assume that the narrowband flux is solely produced by the \lya\ line. 
We do not correct for the continuum, which is extremely weak for
these objects and not always measurable. The mean observed-frame 
Ly$\alpha$ equivalent width (EW) in the stacked normalized
spectra is 600~\AA\  at $z=6.6$ and 410~\AA\ at $z=5.7$, while our
selection chooses LAEs with observed-frame \lya\ EWs
of 130~\AA\ or greater. Even for these lowest EW
objects, the continuum correction to the luminosities is 
$<0.1$ dex (\citealt{taylor21}).
We use aperture-corrected $2''$ diameter aperture magnitudes, with the correction
($\sim 0.30$~mag) determined by comparing with $4''$ diameter apertures. 
We next divide the fluxes corresponding to the narrowband magnitudes
by the filter transmission efficiencies at the spectroscopic redshifts to obtain our final \lya\
line fluxes. We then determine the \lya\ luminosities, \llya, using the luminosity distances
for our adopted cosmological geometry.

As a check, we compare our narrowband-derived \llya\
to our measured $A\times$FWHM, 
which is roughly the spectral line strength. As can be seen from Figure~\ref{ampl_test}, 
there is a near-linear relation between the two, which gives us confidence in our \llya\ 
measurements.

We plan on releasing an atlas of our LAE spectra on a regular basis as
we update the sample. Our goal is to provide all
of the information needed to underpin theoretical modeling.
For each of the two redshifts, $z=5.7$ and $z=6.6$,
we will release a binary FITS file containing the 
quantities given in Table~\ref{Atlas}.
Table~\ref{Atlas}  gives the first release as part of the present paper.

\begin{figure*}
\centerline
{\hskip -1.0cm
\includegraphics[width=1.15\columnwidth]{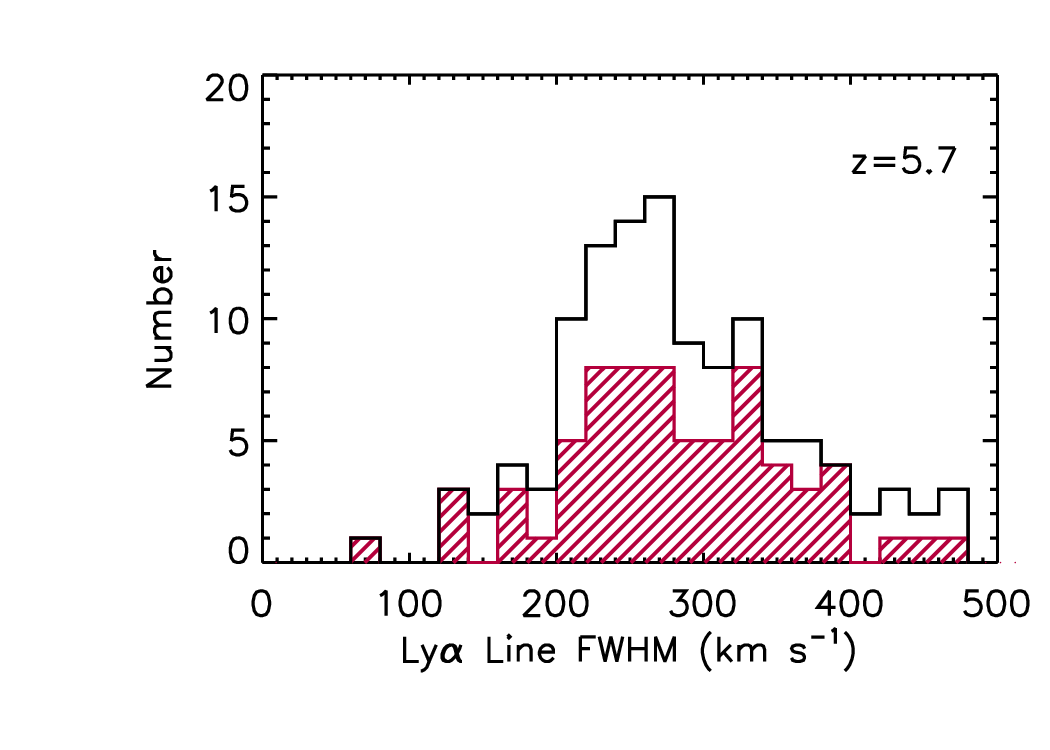}
\hskip -1.0cm
\includegraphics[width=1.15\columnwidth]{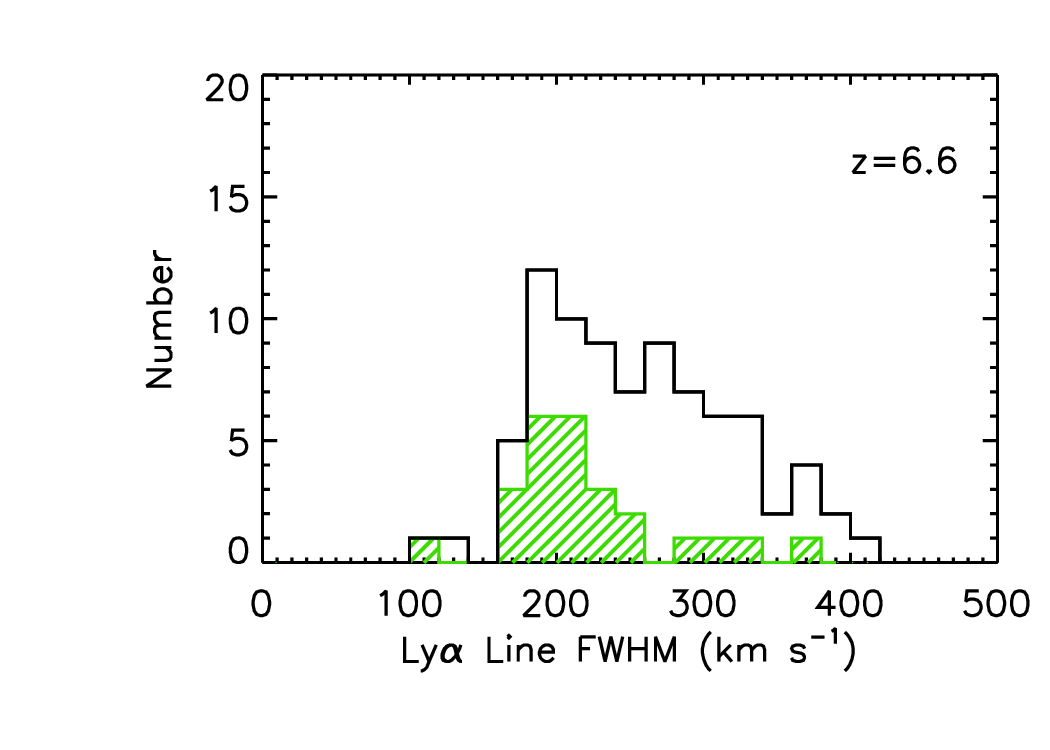}
}
\vskip -0.5cm
\caption{
\lya\ line FWHM distributions for (left) the $z=5.7$ LAEs, where the red
shading shows the lower luminosity sources
($\log$ L(\lya)$<43.25$~erg~s$^{-1}$), and
(right) the $z=6.6$ LAEs, where the green
shading shows the lower luminosity sources.
}
\label{wid_hist}
\end{figure*}

\begin{figure*}
\centerline{
\includegraphics[width=0.5\linewidth]{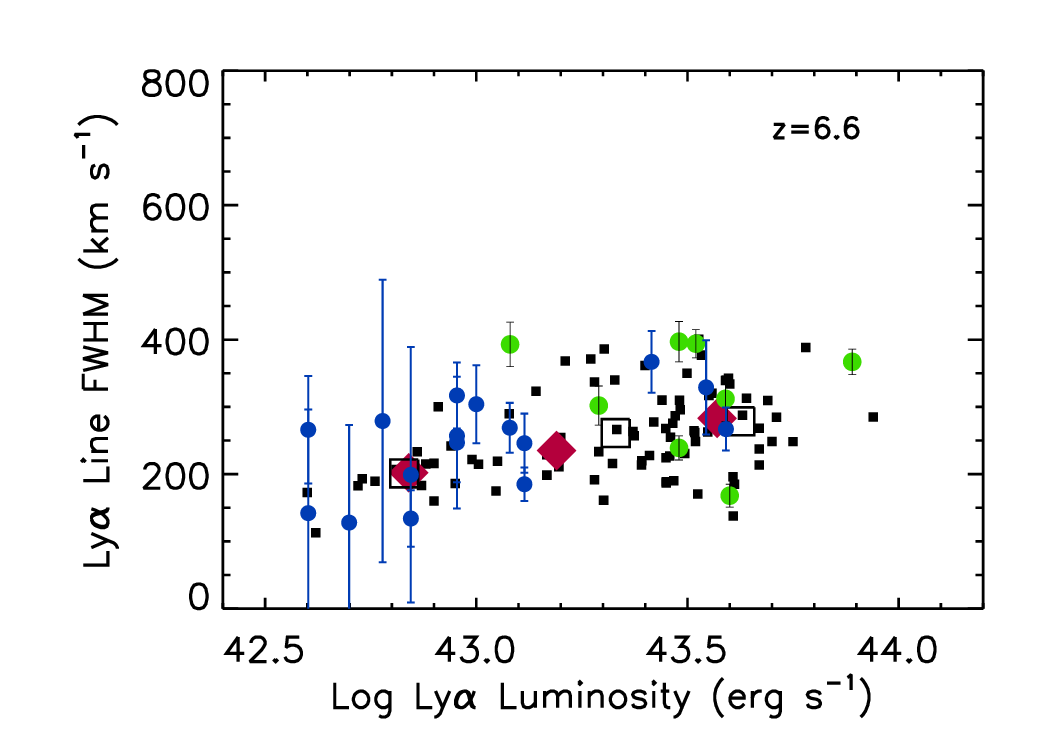}
\includegraphics[width=0.5\linewidth]{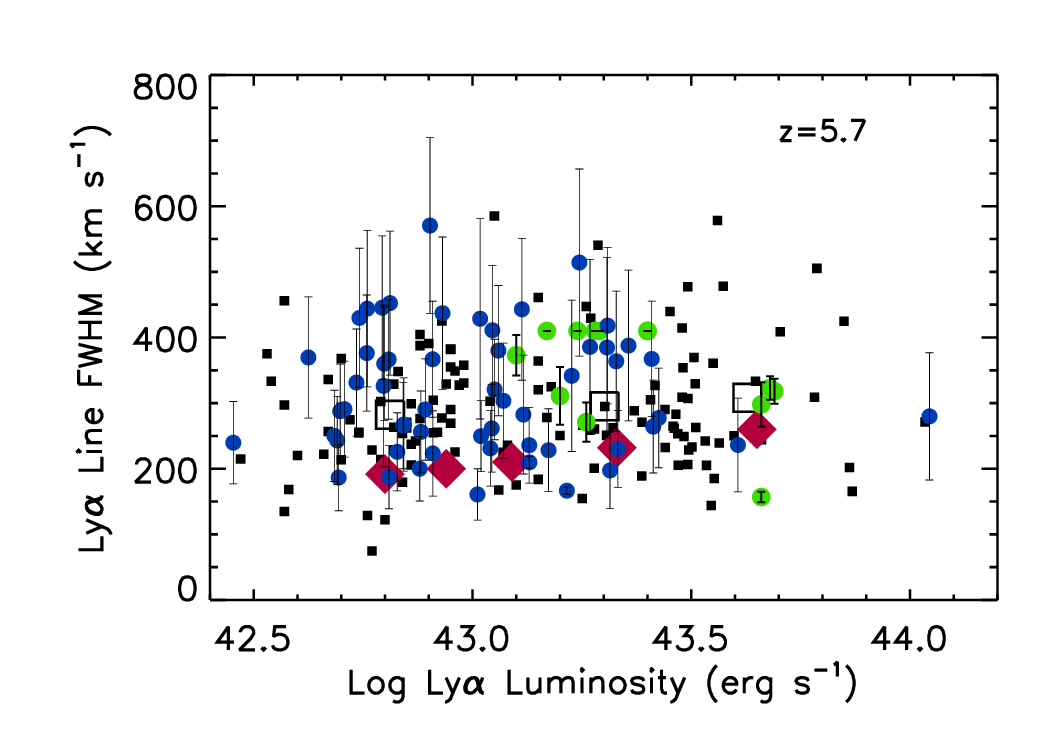}
}
\caption{Comparison of our (left) $z=6.6$ and (right) $z=5.7$ \lya\ line widths (small black squares) 
and their means (large open squares) 
with archival data (left: green circles---\citealt{shibuya18}; blue circles---\citealt{ouchi10};
right: green circles---\citealt{shibuya18}; blue circles---\citealt{mallery12}). In both panels, 
we also show the stacked measurements from \citet{ning22} (red diamonds). 
For clarity, we do not show the errors on the present data 
here, but they are shown in Figure~\ref{LF_LW}. The errors shown on the archival data 
are 68\% confidence on the FWHM. 
\label{wid_compare}}
\end{figure*}

\section{Evolution of the \lya\ Line Widths with Redshift}
\label{vel_wid}
In Figure~\ref{LF_LW}, we show our \lya\ line widths versus \llya\
(blue circles: $z=5.7$;
red squares: $z=6.6$).  As we discuss
below, the $z=5.7$ sample FWHM distribution is relatively
uniform over the observed luminosity
range. In contrast, the $z=6.6$ sample FWHM distribution 
is much narrower at lower luminosities ($\log$~L(\lya)~$<43.25$~erg~s$^{-1}$)
than at high luminosities. At high luminosities, the two redshift samples have
similar FWHM distributions.

In Figure~\ref{wid_hist}, we compare the 
FWHM distributions of the full samples (black lines) 
with those at lower luminosities
(shaded regions).  At $z=5.7$, the two distributions
are similar, and a Kolmogorov-Smirnov test gives a 0.84 probability
that they could be consistent. At $z=6.6$, the two distributions
are quite different, with the lower luminosity LAEs being
narrower. A Kolmogorov-Smirnov test gives only a $4.7 \times 10^{-5}$
probability that they could be drawn from the same distribution.
The high-luminosity LAEs at $z=6.6$ are consistent with both the 
high and low luminosity LAEs
at $z=5.7$. Comparing the high-luminosity
LAEs at the two redshifts gives  a Kolmogorov-Smirnov probability of 0.51.
(See also the stacked spectra of Figure~\ref{stacks}.)

These results suggest that at $z=6.6$, the high-luminosity LAEs may 
preferentially lie in more highly ionized regions than the lower luminosity
LAEs, while by  $z=5.7$,
the IGM is much more uniformly ionized. We return 
to this point in the discussion.

Our \lya\ line widths versus \llya\ are mostly consistent
with previously published values. In Figure~\ref{wid_compare},
we compare our measurements with the measurements from \citet{shibuya18}
(green circles). We also compare with
\citet{ouchi10} for fainter LAEs at $z=6.6$ and with
\citet{mallery12} for LAEs at $z=5.7$. We do not
compare with \citet{ouchi10} for LAEs at $z=5.7$,
as their errors are too large to be useful. We note
that while their $z=6.7$ errors are also large, within
these errors, their data are consistent with the present
results. Finally, we compare with the stacked values of \citet{ning22}
(red diamonds), which are in very good agreement with
our measurements at $z=6.6$ but appear to be low at $z=5.7$.

\section{Discussion}
\label{discuss}

\subsection{Velocity Widths}
\label{velwid}
The differential evolution of the velocity widths with redshift discussed
in Section~\ref{obs} and shown in Figures~\ref{LF_LW} and
\ref{wid_hist} implies
that the $z=5.7$ LAEs must lie in a relatively uniformly ionized
background, regardless of \llya, while at $z=6.6$,
the high-luminosity LAEs continue to lie in ionized regions, but
the lower luminosity sources show the effects of increasing IGM neutrality. Thus,
at $z=6.6$, it appears that the high-luminosity LAEs either are
producing large ionized bubbles themselves, or they are marking overdense galaxy
sites that are producing such bubbles, while the lower luminosity
LAEs are not. In order to avoid suppressing the red wings of the LAEs with the
radiation damping wings from surrounding neutral hydrogen in the IGM,
we require a surrounding highly ionized region with a size $R$
greater than about 1~pMpc (e.g., \citealt{mason20} and references therein).
This size does not depend on the LAE properties but only on the IGM neutrality.

The requirement that the observed LAEs lie 
in ionized bubbles has been used to estimate lower
limits on the fractional ionized volume in the IGM
(\citealt{malhotra06}). Here we use the present data to
refine this calculation at $z = 6.6$.

We adopt the radius versus luminosity shape dependence from \citet{yajima18},
whose normalization agrees closely with the present work.
We then take the present results 
as an empirical normalization of the $R$--\llya\ relation. Using only the shape
of the Yajima curve and normalizing to 1~pMpc at our transition  
$\log$~\llya\ = 43.25~erg~s$^{-1}$, we obtain
\begin{equation}\label{eq_yajima}
        R({\rm in\,\, pMpc}) = 0.82\ L_{43} ^{0.285} \,,
\end{equation}
where $L_{43}$ is \llya\ in units of $10^{43}$~erg~s$^{-1}$.
\lya\ sources fainter then $\log$ \llya\ = 43.25~erg~s$^{-1}$ have  
bubble sizes smaller than 1 pMpc and  will suffer scattering in the red wing.

\begin{figure}
\centerline{
\includegraphics[width=9.0cm]{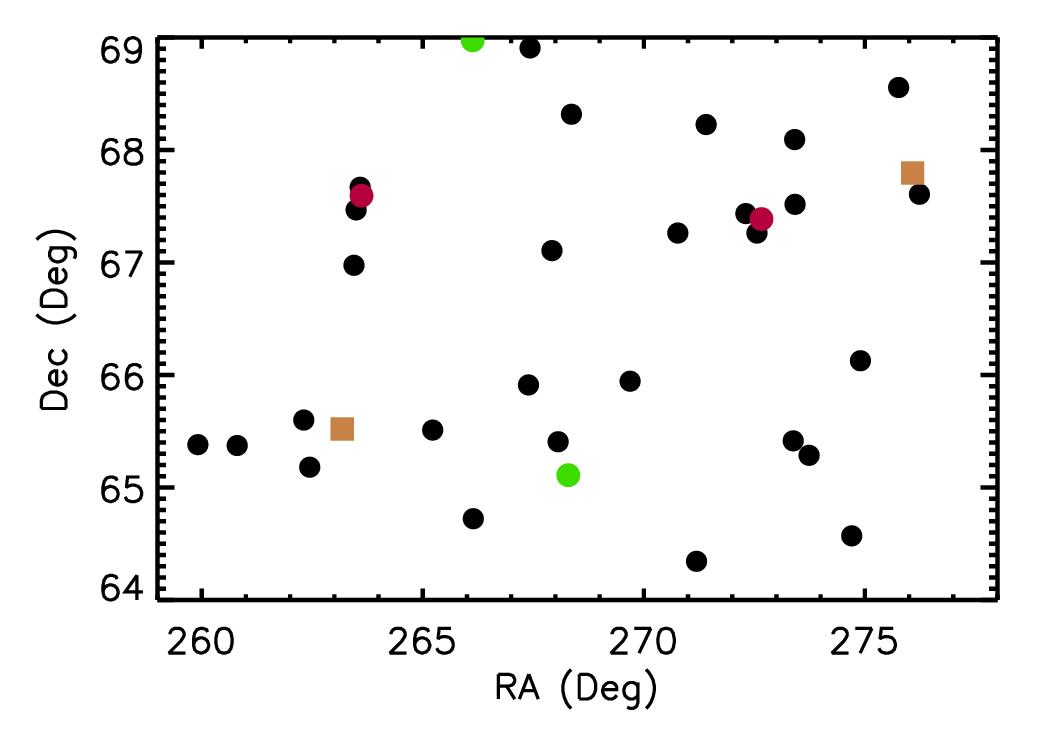}
}
\caption{
Positions of the 32 $z=6.6$ LAEs detected in HEROES 
with $\log$ \llya $>43.4$~erg~s$^{-1}$.
The two sources with double-peaked \lya\ profiles are shown
in green, sources with multiple neighbors within $10'$
are shown in red, and the two broad-line AGNs
are shown as gold squares.
The angular sizes of the bubbles are roughly comparable to the
sizes of the symbols.
\label{z6_groups}
}
\end{figure}

We can now use this calibration of $R$ to estimate the fractional
ionized volume marked by the luminous LAEs. The dependence of the ionized 
volume around a given LAE is
$\propto$ \llya$^{0.855}$, so the ionized fraction
is dominated by the more numerous lower luminosity sources. 
In HEROES, there are 21 sources with
$\log$ \llya\ above 43.5~erg~s$^{-1}$, where the sample is substantially complete
(\citealt{taylor21} give 0.6 for the incompleteness at this luminosity). 
At $\log$ \llya\ above 43.4~erg~s$^{-1}$ there are 32 sources in the field
as shown in Figure~\ref{z6_groups}.
The small covering area of these very luminous sources can be seen in 
Figure~\ref{z6_groups}.
The observed volume is 82000~pMpc$^{3}$,
while the summed incompleteness corrected ionized volume for the 
sources with $\log$ \llya\ above 43.5~erg~s$^{-1}$ is 190~pMpc$^{3}$, or 0.2\% of the volume at $z=6.6$.

Because of the shallow dependence of the volume versus luminosity 
relation of Equation~\ref{eq_yajima},
even lower luminosity sources only mark a small fractional ionized volume. 
We may generalize and extend to lower luminosities using the Schechter function
fit to the $z=6.6$ LAE LF of \citet{ning22} ($\alpha=-1.5$, $\log_{10} \phi _\star$ $=-4.26$,
and $\log_{10}$L$_\star=43.02$, where the units are cMpc$^{-3}$ and erg~s$^{-1}$), we find that ionized
regions for LAEs with $\log$ \llya\ $> 42$~erg~s$^{-1}$ only correspond to about 8\% of the volume. 
This fraction is strictly a lower bound, since it omits yet lower luminosity LAEs and
also ionized regions that do not contain LAEs.

These values are lower than previous estimates made using 
required ionized bubble sizes at these redshifts. 
For example, \citet{malhotra06}
suggested an ionized bubble volume fraction of 20\% to 50\% at $z=6.5$. 
The primary difference from their work is
the adopted LAE LF. \citet{ning22} show a low normalization for the $z=6.6$
LF, in good agreement with \citet{hu10} and up to a factor of 3 lower
than some other measurements of the $z=6.6$ LF. 
These higher normalizations would raise the fraction closer
to the \citet{malhotra06} values. However, \citet{ning20} attribute
such high-normalization LFs to the use of photometric rather than
spectroscopic samples, or to field-to-field variation.
The dependence on the LF does emphasize the uncertainty
of the calculation and that our 8\% ionized volume may be a lower
limit for this reason also.

\subsection{Double-peaked LAEs}
Double-peaked LAEs provide an alternative diagnostic of the
properties of the highly ionized regions surrounding the LAEs. 
The conditions for
seeing double peaks are more restrictive than for seeing
red wings (\citealt{mason20}), since
the blue side of the HII region also needs to be transmissive,
at least to the velocities seen in the blue wing.
\citet{mason20} compute
the proximity radius, $R_{a}$, where the ionizing flux
of the galaxy is sufficient to make the surroundings
have a low enough neutral fraction to pass the blue
light. This radius then determines the most negative
blue velocity that can be seen. From their work, $R_a$ is generally
several times smaller than the radius of the ionized bubble.

At $z\sim3$, $\sim30$\%
of LAEs are double peaked (\citealt{kulas12}). At very low
redshifts, a large fraction of green pea samples (e.g., \citealt{henry15,yang17})
show blue side emission, with $\sim30$\% showing
strong blue side peaks (i.e., the blue peak value exceeds 25\% 
of the red peak value).

\begin{figure}[ht]
{
\includegraphics[height=0.52\columnwidth,width=0.99\columnwidth]{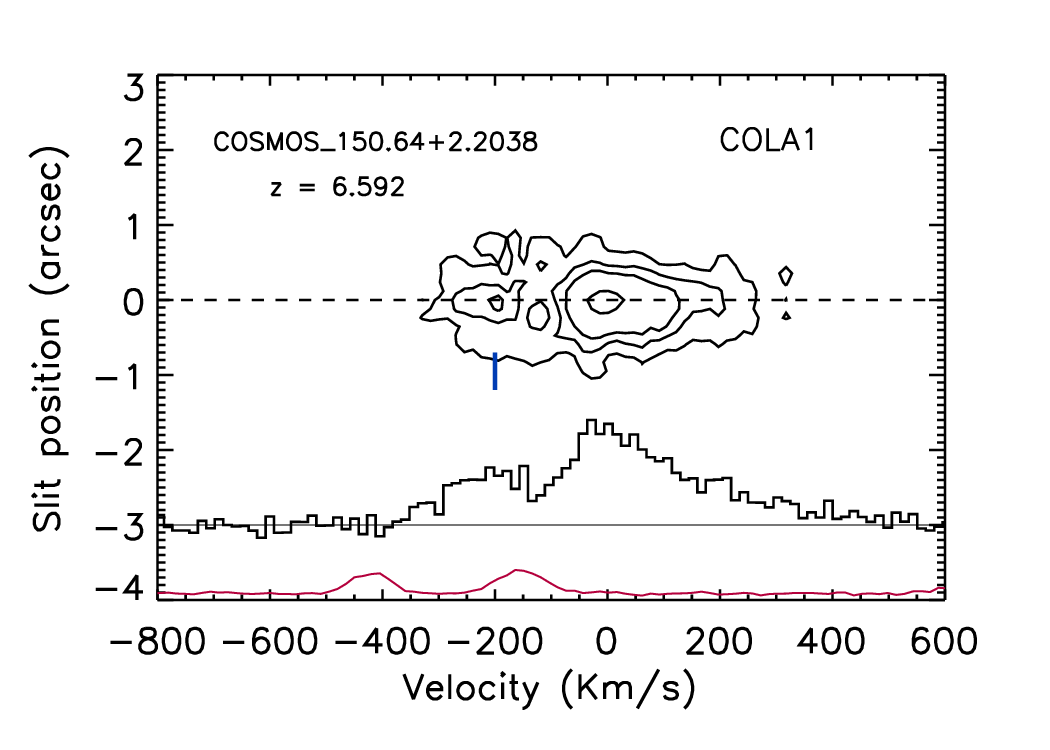}
\includegraphics[height=0.52\columnwidth,width=0.99\columnwidth]{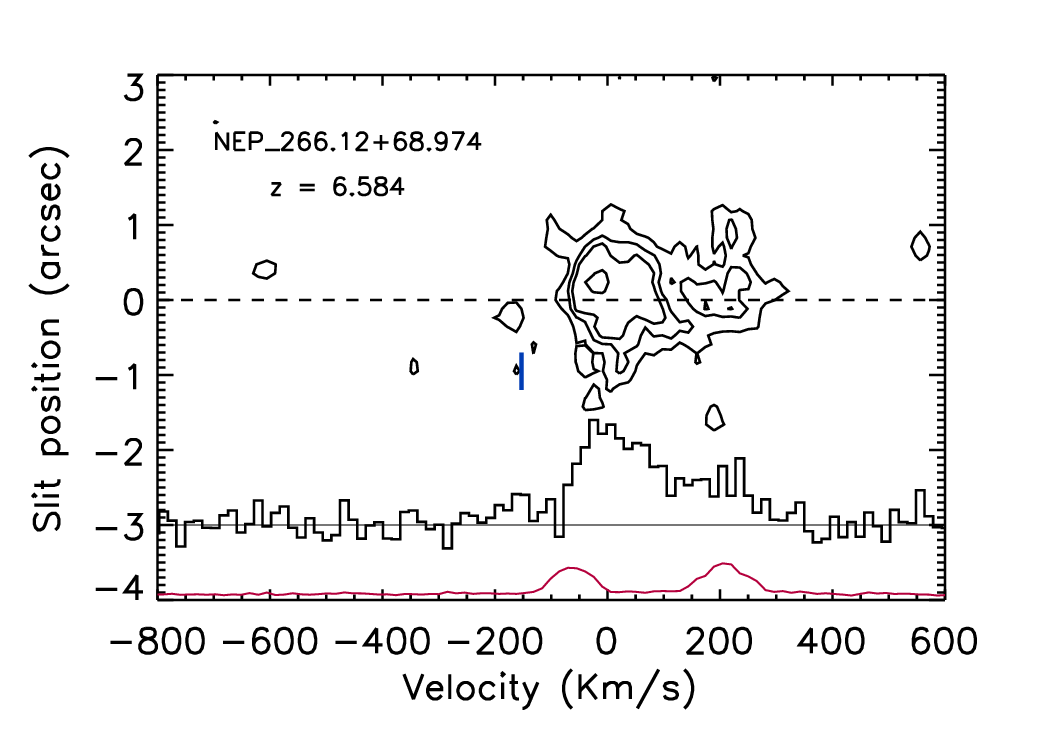}
\includegraphics[height=0.52\columnwidth,width=0.99\columnwidth]{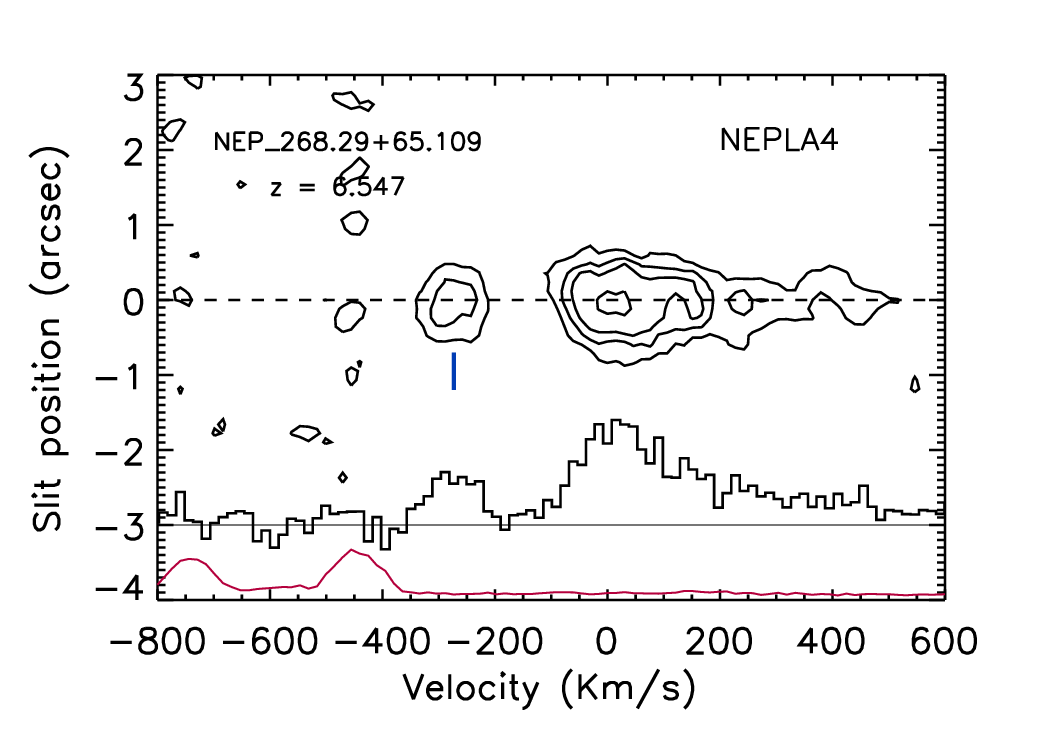}
\includegraphics[height=0.52\columnwidth,width=0.99\columnwidth]{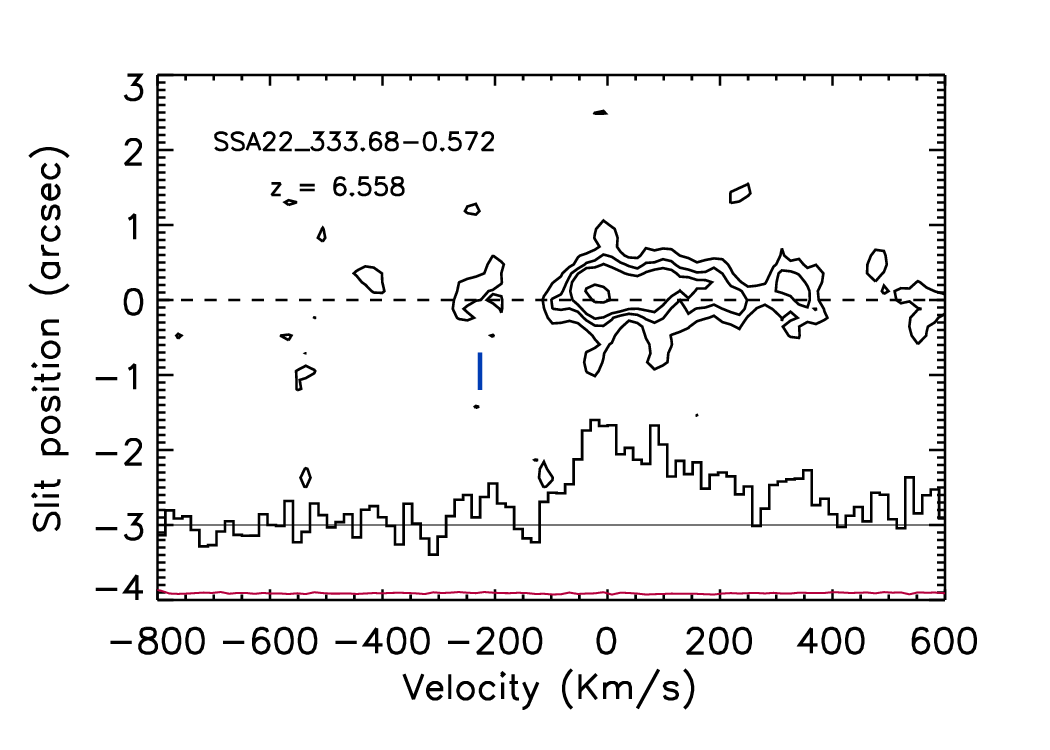}
}
\caption{
2D spectral images of the four double-peaked sources in the $z=6.6$ sample. 
The contour levels
are 0.2, 0.36, 0.5, and 0.9 times the red peak value. 
We previously published two of these (COLA1 and NEPLA4), which we label.
The dashed line marks
the slit center, and the small blue vertical line the second peak.  
The 1D spectrum is also shown, together with
the sky spectrum (red), both of which have arbitrary normalizations to fit
on the plot.
\label{double_z6}
}
\end{figure}

\begin{figure}[ht]
{
\includegraphics[height=0.52\columnwidth,width=0.99\columnwidth]{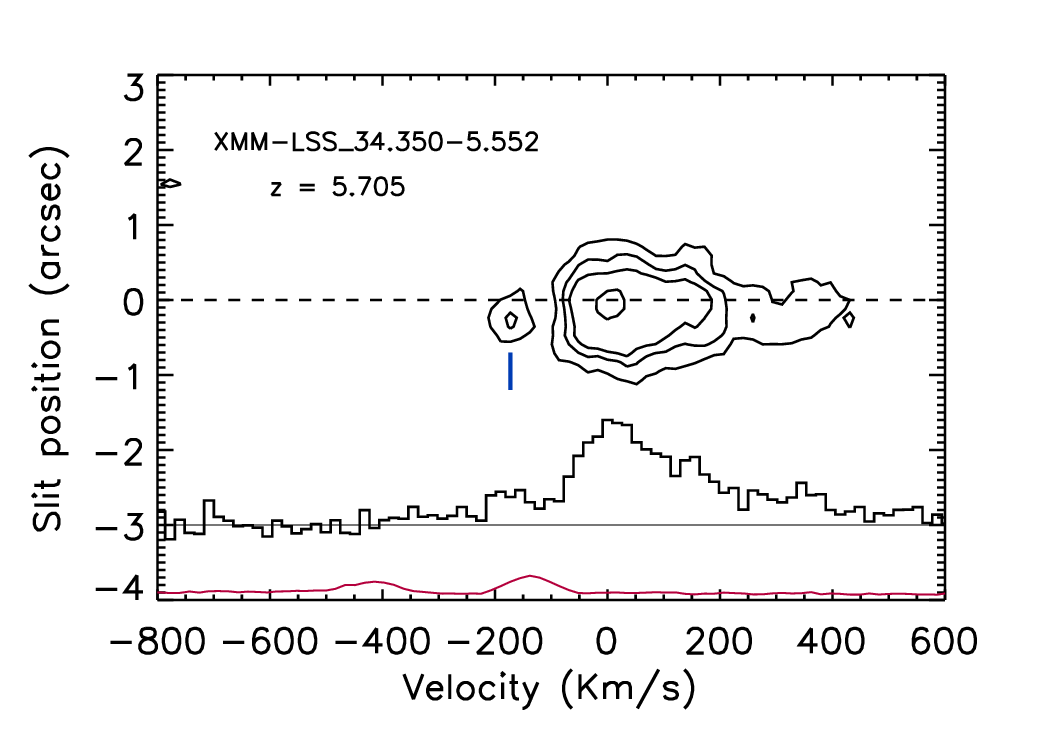}
\includegraphics[height=0.52\columnwidth,width=0.99\columnwidth]{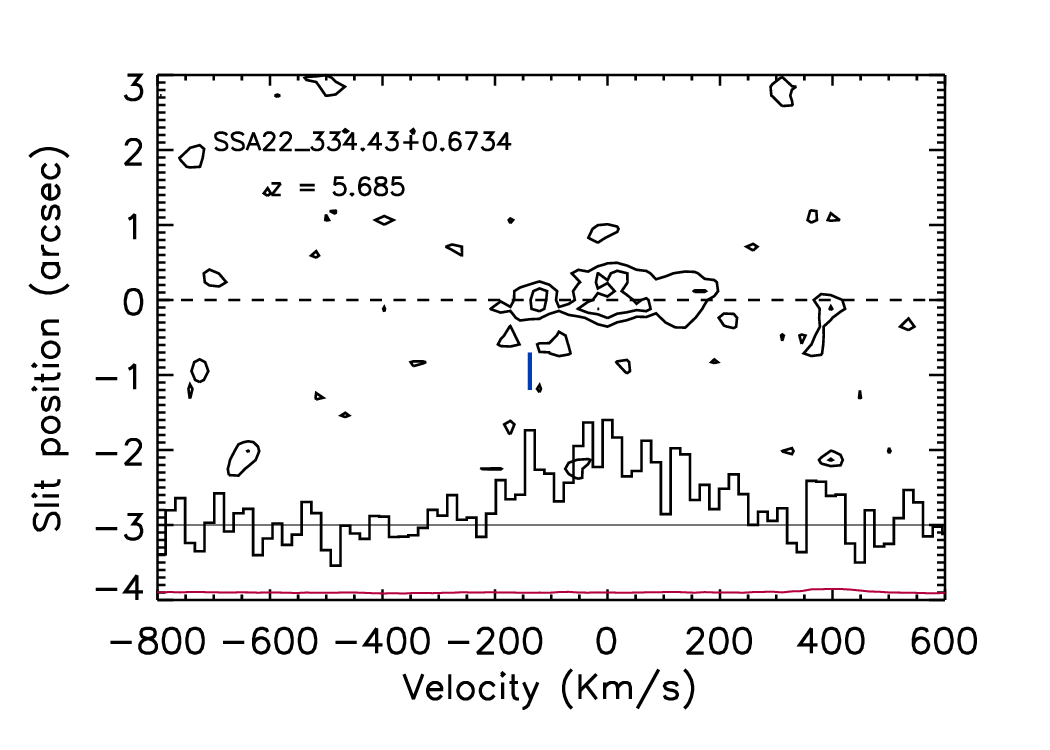}
\includegraphics[height=0.52\columnwidth,width=0.99\columnwidth]{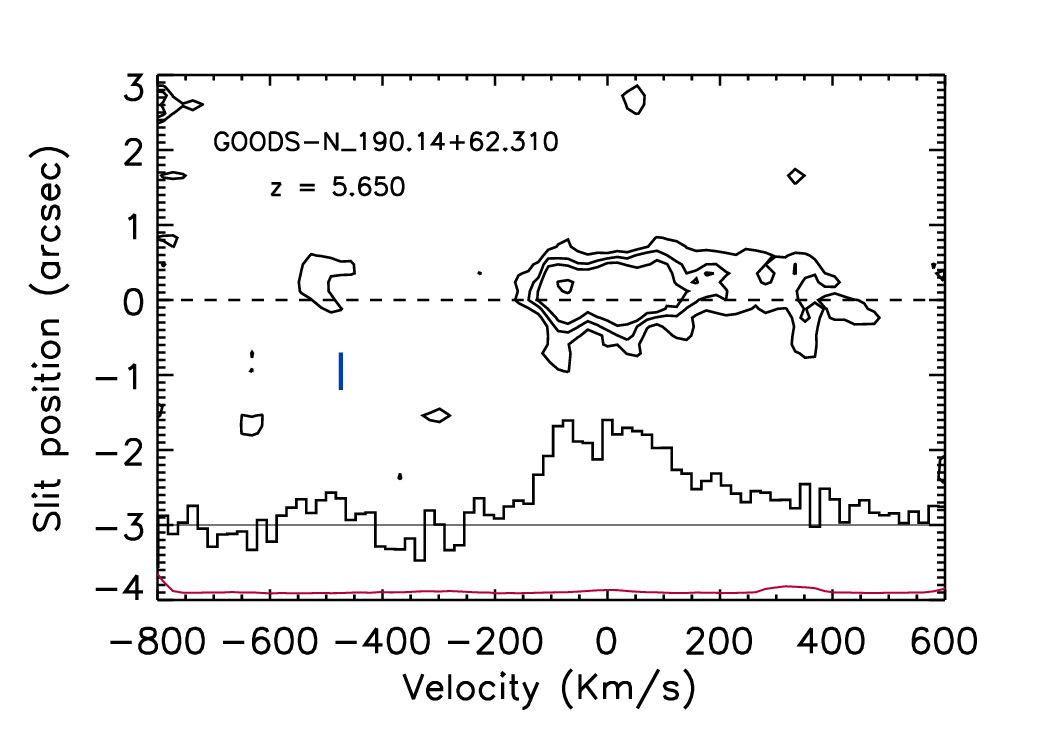}
\includegraphics[height=0.52\columnwidth,width=0.99\columnwidth]{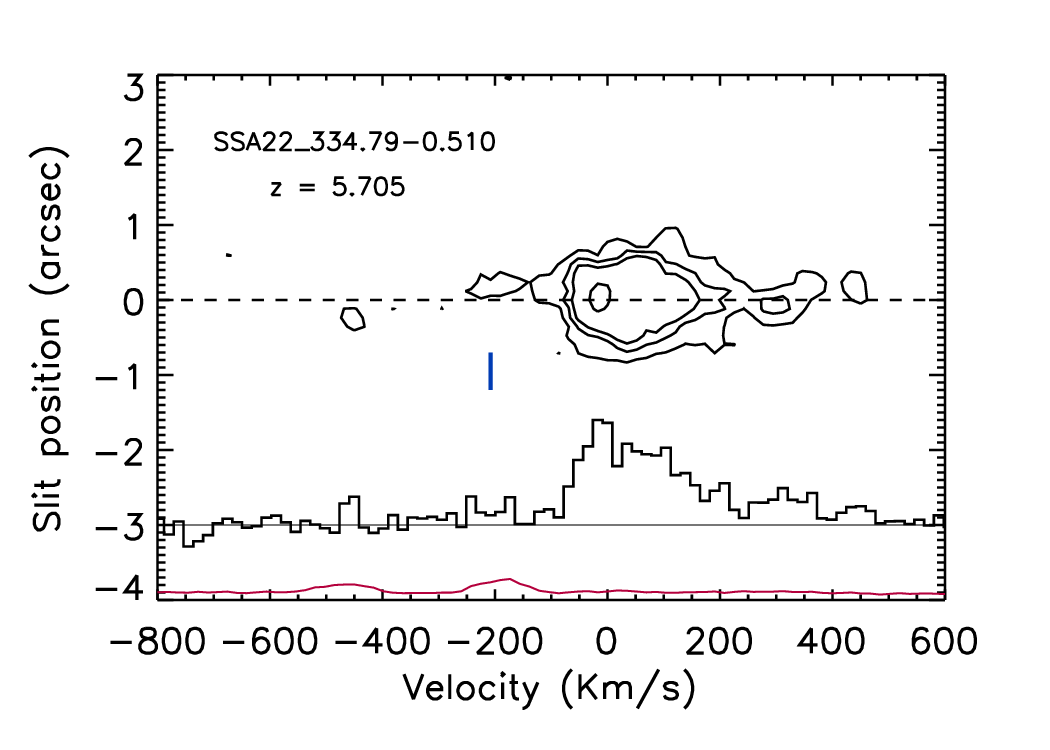}
}
\caption{
2D spectral images of four of the fifteen double-peaked sources in the $z=5.7$ sample.
The contour levels are 0.2, 0.36, 0.5, and 0.9 times the red peak value. The dashed line marks
the slit center, and the small blue vertical line the second peak. 
The 1D spectrum is also shown, together with
the sky spectrum (red), both of which have arbitrary normalizations to fit on the plot.
\label{double_z5}
}
\end{figure}

\begin{figure}
\centerline{
\includegraphics[width=9.0cm]{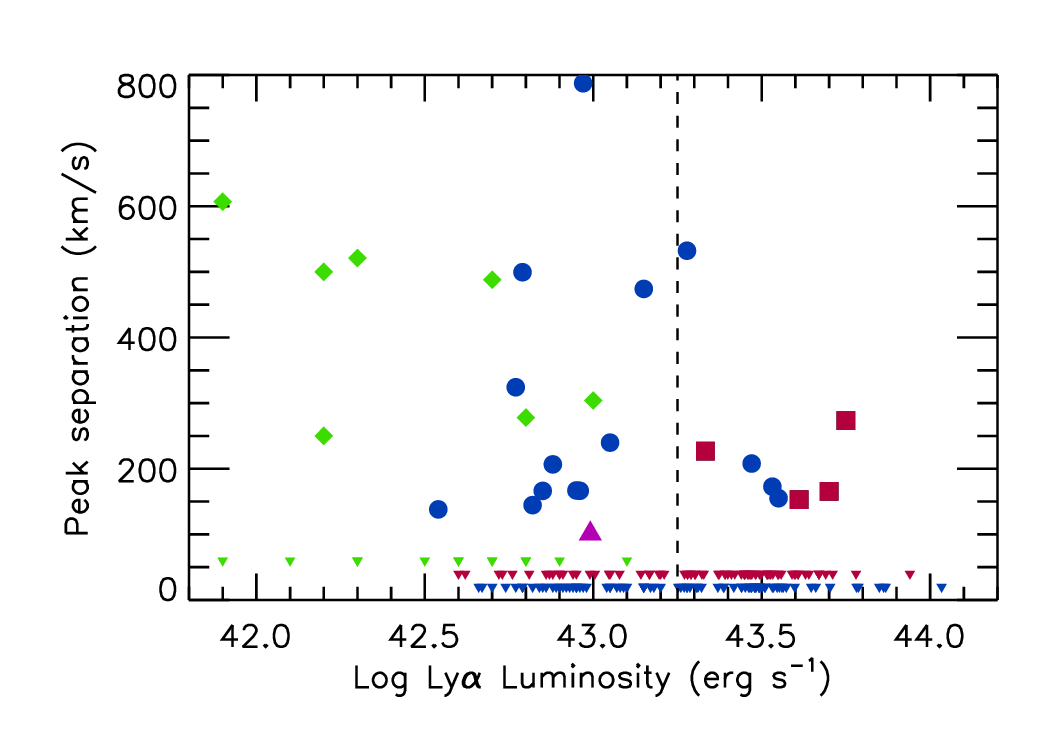}
}
\caption{Peak velocity separation vs. $\log$ \llya\ for the $z=6.6$ double-peaked LAEs
(red squares) and the $z=5.7$ double-peaked LAEs (blue circles). 
We also show the double-peaked sources in the low-redshift Green Pea galaxy sample
of \citet{yang17} (green diamonds) 
and the $z=6.9$ double-peaked LAE
of \citet{meyer21} (purple triangle). Sources without double peaks are marked
with small triangles  at the bottom of the figure (red is $z=6.6$, blue is $z=5.7$).
The Green Pea sources which do not satisfy our double peak definition are
shown as green triangles.
\label{compare_peaks}
}
\end{figure}

Here we quantify the double-peaked fractions at $z=5.7$
and $z=6.6$ to compare with each other and with the low-redshift
fraction. We restrict to high-quality spectra, where the signal-to-noise (S/N) 
of the amplitude in the
spectral fits exceeds 7. To make an objective selection, we take the redshift
of the main peak and blank the 2D spectral image redward of
$-120$~km~s$^{-1}$. This choice minimizes contamination from the
blue tail of the main peak, given the resolution of the spectroscopic observations. 
We then search the central region
of the blanked 2D image ($\pm0\farcs6$ from the slit position of the main peak)
for secondary
peaks, requiring that the S/N in the secondary peak lies
above $4\sigma$ and that the secondary peak value exceeds 
25\% of the main peak value.
Finally, we restrict to secondary
peaks with velocities redward of $-800$~km~s$^{-1}$ relative
to the main  peak velocity. Our final sample therefore corresponds
to source with double peak separations between 120 and 800~km~s$^{-1}$.
After visual inspection, we decided that apparent double peaks in two of the
$z=5.7$ LAEs were artifacts; thus, hereafter, we consider them as single-peaked LAEs.

At $z=6.6$, four of the 71 high-quality LAEs that we searched are double-peaked
sources. We show these in Figure~\ref{double_z6}
(two of the four are our previously published sources COLA1 and NEPLA4,
while the other two are more marginal detections).
At $z=5.7$, 15 of the 107 high-quality LAEs that we searched are double-peaked sources.
We show four examples in Figure~\ref{double_z5}. We mark the double-peaked sources
using flag = 2 in Table~3.

In Figure~\ref{compare_peaks}, we
plot the peak separations of our double-peaked LAEs versus $\log$~\llya. 
We also show the low-redshift Green Pea galaxy sample of \citet{yang17} and the
$z=6.9$ double-peaked LAE of \citet{meyer21}, where this information is available.
We do not have such quantitative information for the intermediate redshift samples.

At $z=5.7$, the fraction of double-peaked LAEs at $\log$~\llya$>43.25$~erg~s$^{-1}$
is consistent with the overall fraction at this redshift ($14\%\pm4$\%).
At $z=6.6$, the fraction of double-peaked LAEs 
is $6 (3-10)$\%. We do not see any double peaks in lower luminosity
sources at $z=6.6$,  though we note that the $z=6.9$ double-peaked
LAE of \citet{meyer21} lies at lower luminosity.
At both redshifts, the fraction of double-peaked LAEs is low compared to
the 30\% seen at $z \lesssim 3$. 
This is consistent with our expectation that the blue-side
scattering by neutral material in the ionized regions
may be obscuring the blue peaks in a significant fraction
of the sources at these redshifts, despite their having substantially
unobscured red-side emission. 

\begin{figure}
\centerline{
\includegraphics[width=9.5cm]{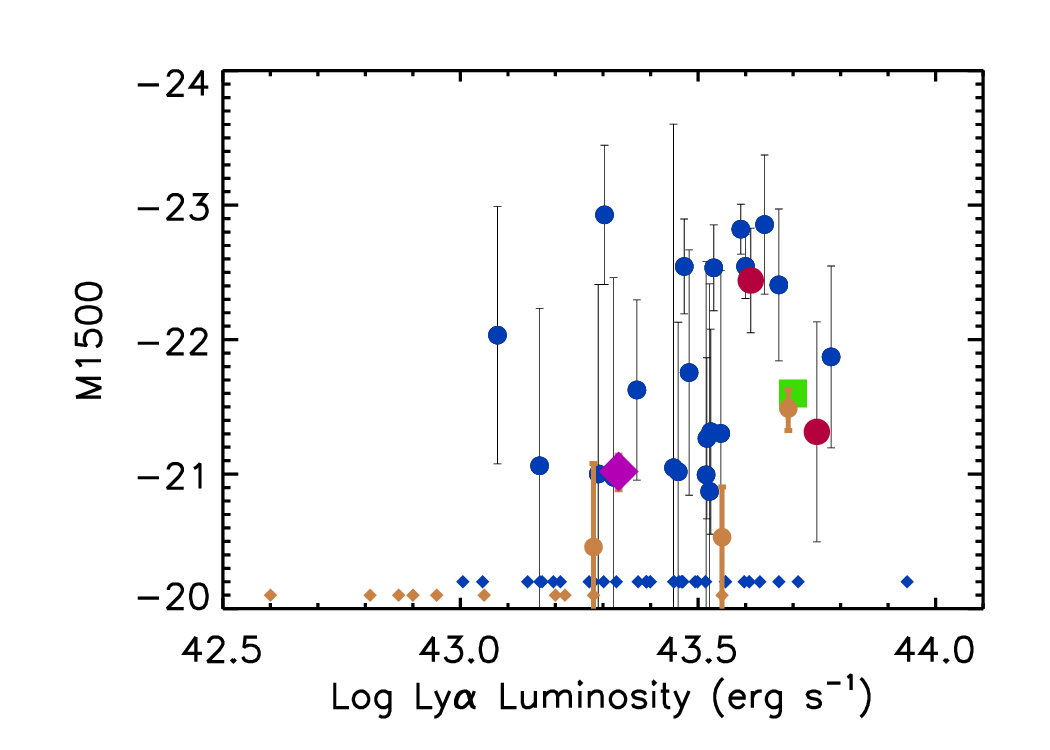}
}
\caption{UV absolute magnitude at 1500~\AA\ vs. $\log$ \llya\ for the $z=6.6$ LAEs
in HEROES (blue circles) and SSA22 (gold circles). 
The $Y$ magnitudes used to calculate M1500 in HEROES are the Kron magnitudes from \citet{taylor23}.
Detected sources are shown with $1\sigma$ error bars. Undetected sources
are shown as small diamonds  at nominal values of -21.8 (HEROES) and -21.9 (SSA22). 
The two double-peaked LAEs
in HEROES are shown as red circles, and the one in SSA22 as a purple diamond. 
COLA1 (green square) is shown
using the M1500 value from \citet{matthee18}.
\label{la_nuv}
}
\end{figure}

\begin{figure}
\centerline{
\includegraphics[width=9.5cm]{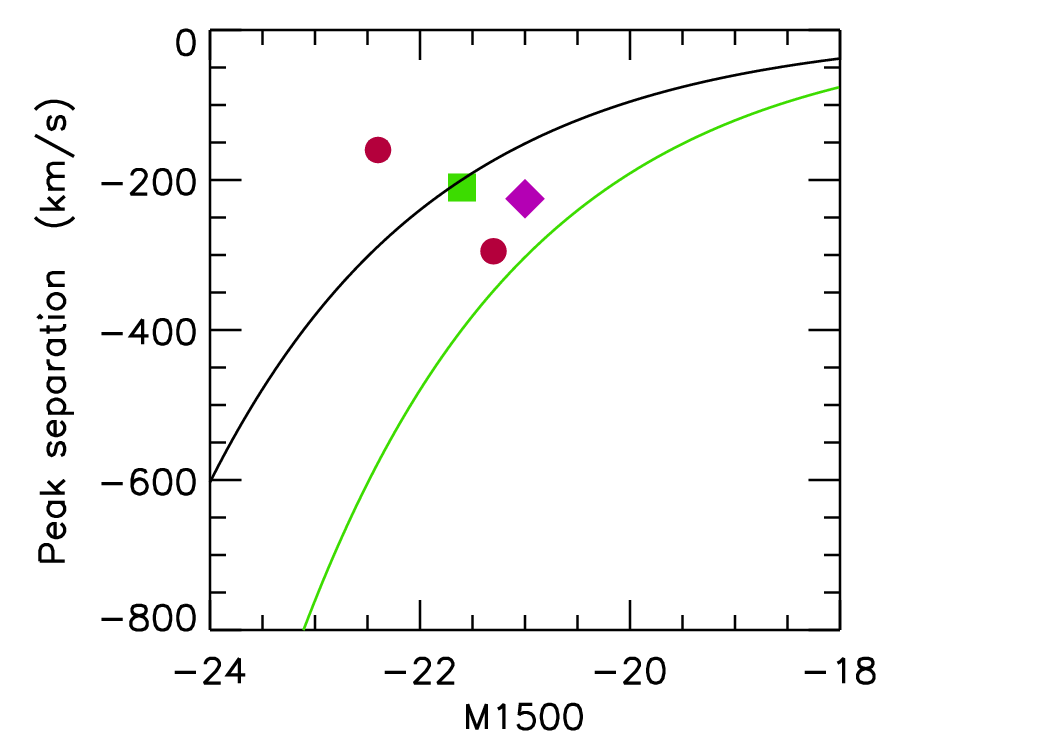}
}
\caption{Peak velocity separation vs. UV absolute magnitude at 1500~\AA\
for the $z=6.6$ double-peaked LAEs (red circles---HEROES;
green square---COLA1; purple diamond---SSA22-333.68-0.572).
The black curve shows the reference model of \citet{mason20},
while the green curve shows the same model but with half the average
density in the neighborhood of the source.
\label{charlotte_plot}
}
\end{figure}

The proximity radius is larger for sources with brighter near-UV (NUV)
luminosities, so we might expect the double-peaked LAEs preferentially
to have higher 1500~\AA\ absolute magnitudes (M1500). Indeed, all of the 
double-peaked LAEs have detected $Y$ magnitudes, which we have used to
compute M1500 following \citet{matthee18}, though NEPLA4 is only
detected at the $2\sigma$ level. We do not attempt to correct
for the contributions of the \lya\ line in the wing of the $Y$ band,
which would only make the M1500 fainter.
(The error bars for all the sources are shown in Figure~\ref{la_nuv}.)
In Figure~\ref{la_nuv}, we compare the double-peaked LAEs
with other LAEs in the HEROES and SSA22 fields.
We can see that the double-peaked LAEs (larger symbols) are among the 
brighter NUV sources, with M1500 magnitudes in the range from -21 to -22.5.

\citet{mason20} computed models for the proximity radius
at $z=6.6$ and the corresponding peak velocity separation as a function of the NUV luminosity. 
In Figure~\ref{charlotte_plot}, we show that three of our four $z=6.6$ 
double-peaked LAEs lie beneath the \citet{mason20}
reference model (black curve) for the peak velocity separation
and are therefore  inconsistent with this model
(as \citealt{mason20} pointed out for COLA1).  However, the \citet{mason20}
models show that reducing the
local gas density by a factor of 2 (green curve) would
permit the transmission of the blue peaks. 

As \citet{mason20} point out, their reference model is highly optimistic,
assuming an escape fraction of one and no clumping, so even
the green curve may not be sufficient. 
For example, \cite{meyer21}
argue by comparing the peak velocity separations of such wide sources 
with those of local peak separations
that the escape fractions should be considerably lower, with values
of around 0.1--0.2 (\citealt{izotov18}).
Thus, it may require multiple
ionizing sources in the vicinity of the double-peaked LAEs
to generate the large proximity radii necessary
to see the blue peaks in these sources. Deeper observations in both the continuum
and \lya\ in the vicinity of the double-peaked LAEs are clearly needed to
examine this possibility.

\section{Summary}
\label{summary}

We present two uniformly observed large spectroscopic samples of LAEs
(127 at $z=5.7$ and 82 at $z=6.6$) from our Keck/DEIMOS follow-up of
narrowband selected sources in the large, contiguous HEROES field (44~deg$^2$), plus several 
smaller fields from the Subaru/HSC and Subaru/Suprime-Cam archives.
Our main results are as follows:

The generic spectral shape of the average \lya\ line profile at these redshifts,
with a sharp break at the blue side and a tail to the red side, belies the diversity
of individual line shapes and widths that we see. This emphasizes the need for large 
spectroscopic samples, such as the present atlas, which   
can be combined with models of the 
IGM density and ionization structure to predict the LAE properties at higher redshifts.

Large spectroscopic samples, such as the present atlas, also reveal rare and interesting
objects.  We found small numbers of LAEs with double-peaked \lya\ profiles; very
extended red wings;  broad-line AGNs; and a unique 
lensed LAE cross. 

By comparing the \lya\ line widths for the LAEs at the two redshifts, 
we found that the FWHM distribution for the $z=5.7$ sample is
relatively uniform 
over the observed luminosity range, while that for the $z=6.6$ sample is 
not:  it is much narrower at lower luminosities 
($\log$~L(\lya)~$<43.25$~erg~s$^{-1}$) than at high luminosities. The 
$z = 5.7$ and $z = 6.6$  samples have similar FWHM 
distributions at high luminosities.
This suggests that at $z = 6.6$, the high-luminosity LAEs may preferentially lie in more 
highly ionized regions than the lower luminosity LAEs. The ionized regions
constitute at least 8\% of the volume at $z = 6.6$, but this estimate
is quite uncertain and depends strongly on the adopted \lya\ LF.

At $z = 5.7$, about 14\% of the LAEs have double peaks, while 
at $z = 6.6$, this has dropped to about 6\%. This can be compared
with the 30\% seen at lower redshifts. Comparing the observed $z = 6.6$
double peaks with the models of \citet{mason20} suggests that the LAEs themselves
are not UV luminous enough to ionize the surrounding regions to the required
level (see also \citealt{meyer21}). The 
double-peaked \lya\ profiles may therefore mark sites with multiple ionizing
sources.

In future work, we plan to migrate our ongoing spectroscopic follow-up from Keck/DEIMOS
to the Keck/KCWI integral field spectrograph, which will provide spatially resolved spectra of the LAEs
and also allow us to search for neighboring ionizing sources.

\begin{acknowledgements}

We gratefully acknowledge support for this research from NSF grants AST-1716093 (A.~S., E.~M.~H.) 
and AST-1715145  (A.~J.~B), NASA grant 80NSSC24K0618, a Kellett Mid-Career Award and a WARF Named Professorship 
from the University of Wisconsin-Madison Office of the Vice Chancellor for
 Research and Graduate Education with funding from the Wisconsin Alumni Research 
Foundation (A.~J.~B.), and the William F. Vilas Estate (A.~J.~T.).

The Hyper Suprime-Cam (HSC) collaboration includes the astronomical communities of Japan and Taiwan, and Princeton University. The HSC instrumentation and software were developed by the National Astronomical Observatory of Japan (NAOJ), the Kavli Institute for the Physics and Mathematics of the Universe (Kavli IPMU), the University of Tokyo, the High Energy Accelerator Research Organization (KEK), the Academia Sinica Institute for Astronomy and Astrophysics in Taiwan (ASIAA), and Princeton University. Funding was contributed by the FIRST program from the Japanese Cabinet Office, the Ministry of Education, Culture, Sports, Science and Technology (MEXT), the Japan Society for the Promotion of Science (JSPS), Japan Science and Technology Agency (JST), the Toray Science Foundation, NAOJ, Kavli IPMU, KEK, ASIAA, and Princeton University.

This paper is based on data collected at the Subaru Telescope and retrieved from the HSC data archive system, 
which is operated by the Subaru Telescope and Astronomy Data Center (ADC) at NAOJ. 
Data analysis was in part carried out with the cooperation of Center for Computational Astrophysics (CfCA), NAOJ. 

This paper makes use of software developed for the Vera C. Rubin Observatory. 
We thank the Rubin Observatory for making their code available as free software. 

We wish to recognize and acknowledge the very significant cultural role and 
reverence that the summit of Maunakea has always had within the 
indigenous Hawaiian community. We are most fortunate
to have the opportunity to conduct observations from this mountain.

\end{acknowledgements}

\facilities{Keck, Subaru}

\startlongtable
\begin{deluxetable*}{ccccccccc}
\tablecaption{$z=5.7$\ Ly$\alpha$ sample  \label{z5tab}}
\tablewidth{350pt}
\tablehead{
\colhead{Field} & \colhead{R.A.} & \colhead{Decl.} &
\colhead{$z$} &
\colhead{$\log {\rm L(Ly}\alpha$)} 
& \colhead{FWHM} & \colhead{error} & \colhead{a$_{asym}$} 
& \colhead{error}\\
\colhead{} &  \colhead{} & \colhead{} & \colhead{} &
\colhead{(erg~s$^{-1}$)}  & \multicolumn{2}{c}{$({\rm km\ s}^{-1})$}
}
\startdata
XMM-LSS  &   
33.86914 & -4.81525 & 5.7166 & 43.48 & 414.61 & 25.47 & 0.26 & 0.02 \\
XMM-LSS  &   
34.10295 & -4.92128 & 5.7050 & 43.41 & 305.15 & 36.75 & 0.10 & 0.06 \\
XMM-LSS  &   
34.35014 & -5.55270 & 5.7060 & 43.53 & 242.08 & 11.97 & 0.31 & 0.01 \\
XMM-LSS  &   
34.65159 & -5.59116 & 5.6973 & 43.39 & 188.88 & 16.98 & 0.41 & 0.02 \\
XMM-LSS  &   
35.35540 & -3.67684 & 5.6982 & 43.56 & 578.22 & 58.60 & 0.46 & 0.04 \\
XMM-LSS  &   
35.35643 & -4.13690 & 5.6760 & 43.47 & 253.17 & 16.14 & 0.26 & 0.02 \\
XMM-LSS  &   
35.94127 & -3.90492 & 5.7360 & 43.28 & 260.58 & 11.62 & 0.16 & 0.02 \\
XMM-LSS  &   
35.49671 & -4.97522 & 5.6968 & 43.37 & 288.28 & 18.04 & 0.08 & 0.03 \\
A370     &   
39.97289 & -1.60773 & 5.6933 & 42.85 & 270.70 & 22.31 & 0.17 & 0.04 \\
A370     &   
40.06517 & -1.49633 & 5.7122 & 42.90 & 390.72 & 38.96 & 0.17 & 0.04 \\
A370     &   
40.12097 & -1.65553 & 5.6437 & 42.95 & 381.85 & 53.90 & 0.06 & 0.07 \\
A370     &   
40.13245 & -1.54178 & 5.7126 & 42.57 & 296.97 & 63.04 & 0.15 & 0.10 \\
A370     &   
40.14673 & -1.60750 & 5.6831 & 42.87 & 242.36 & 46.98 & 0.48 & 0.04 \\
A370     &   
40.16127 & -1.58338 & 5.7080 & 42.77 & 228.42 & 20.07 & 0.27 & 0.03 \\
A370     &   
40.17686 & -1.55455 & 5.6561 & 42.85 & 262.02 & 37.95 & 0.10 & 0.07 \\
A370     &   
40.23523 & -1.47936 & 5.7078 & 42.79 & 302.47 & 25.76 & 0.33 & 0.03 \\
A370     &   
40.24645 & -1.46045 & 5.7300 & 42.89 & 292.69 & 25.77 & 0.31 & 0.03 \\
A370     &   
40.26759 & -1.46400 & 5.7310 & 42.88 & 404.41 & 31.62 & 0.20 & 0.03 \\
A370     &   
40.34016 & -1.53909 & 5.6789 & 42.79 & 214.26 & 20.64 & 0.05 & 0.05 \\
GOODS-N  &   
188.76352 & 62.28720 & 5.6985 & 43.04 & 302.82 & 30.51 & 0.23 & 0.03 \\
GOODS-N  &   
188.81929 & 62.08558 & 5.7220 & 42.77 & 74.86 & 15.23 & 0.38 & 0.03 \\
GOODS-N  &   
188.88393 & 62.24594 & 5.7026 & 42.57 & 134.96 & 47.57 & 0.41 & 0.06 \\
GOODS-N  &   
188.99542 & 62.17142 & 5.6733 & 42.58 & 168.20 & 88.07 & 0.53 & 0.09 \\
GOODS-N  &   
189.03284 & 62.14396 & 5.6406 & 42.95 & 354.87 & 23.31 & 0.15 & 0.03 \\
GOODS-N  &   
189.03906 & 62.04567 & 5.7200 & 42.72 & 274.77 & 46.38 & 0.35 & 0.05 \\
GOODS-N  &   
189.05014 & 62.07235 & 5.7410 & 42.82 & 263.38 & 36.40 & 0.01 & 0.07 \\
GOODS-N  &   
189.05605 & 62.13004 & 5.6350 & 42.95 & 269.44 & 34.37 & 0.31 & 0.04 \\
GOODS-N  &   
189.21527 & 62.32684 & 5.6763 & 42.96 & 349.14 & 33.95 & 0.06 & 0.05 \\
GOODS-N  &   
189.21686 & 62.36461 & 5.6890 & 42.92 & 255.62 & 19.77 & 0.12 & 0.04 \\
GOODS-N  &   
189.32457 & 62.29978 & 5.6630 & 42.91 & 254.76 & 33.77 & 0.31 & 0.04 \\
GOODS-N  &   
189.43633 & 62.19595 & 5.6753 & 43.26 & 447.39 & 57.51 & 0.28 & 0.05 \\
GOODS-N  &   
189.48480 & 62.40435 & 5.6486 & 42.60 & 220.31 & 35.71 & 0.18 & 0.07 \\
GOODS-N  &   
189.57741 & 62.27264 & 5.7275 & 43.55 & 360.75 & 57.98 & 0.18 & 0.07 \\
GOODS-N  &   
189.59084 & 62.17962 & 5.6490 & 42.83 & 348.24 & 74.42 & 0.28 & 0.07 \\
GOODS-N  &   
189.64726 & 62.11198 & 5.6940 & 42.80 & 122.23 & 30.11 & 0.30 & 0.07 \\
GOODS-N  &   
189.76289 & 62.24575 & 5.7374 & 43.07 & 225.27 & 26.14 & 0.32 & 0.03 \\
GOODS-N  &   
189.97046 & 62.17626 & 5.6356 & 42.98 & 330.55 & 33.84 & 0.31 & 0.03 \\
GOODS-N  &   
190.03114 & 62.21266 & 5.7388 & 42.74 & 254.66 & 19.55 & 0.23 & 0.03 \\
GOODS-N  &   
190.06612 & 62.29416 & 5.6945 & 42.57 & 455.95 & 89.96 & 0.14 & 0.09 \\
GOODS-N  &   
190.13959 & 62.31078 & 5.6502 & 43.15 & 320.44 & 29.03 & 0.18 & 0.04 \\
GOODS-N  &   
190.18228 & 62.25968 & 5.7055 & 42.84 & 179.52 & 27.65 & 0.42 & 0.03 \\
GOODS-N  &   
190.26646 & 62.36457 & 5.7437 & 42.94 & 329.14 & 26.64 & 0.11 & 0.04 \\
GOODS-N  &   
190.28117 & 62.38783 & 5.6865 & 42.66 & 222.16 & 22.44 & 0.17 & 0.04 \\
GOODS-N  &   
190.31573 & 62.38276 & 5.7020 & 43.18 & 325.04 & 34.85 & 0.31 & 0.03 \\
GOODS-N  &   
190.35735 & 62.34753 & 5.7154 & 43.17 & 278.10 & 20.62 & 0.36 & 0.02 \\
GOODS-N  &   
190.36993 & 62.33969 & 5.6960 & 43.44 & 290.44 & 5.24 & 0.34 & 0.01 \\
GOODS-N  &   
190.47531 & 62.41094 & 5.7458 & 42.82 & 228.14 & 12.95 & 0.33 & 0.02 \\
SSA17    &   
256.56177 & 43.80437 & 5.6655 & 42.70 & 367.82 & 81.64 & 0.02 & 0.11 \\
SSA17    &   
256.69565 & 43.75570 & 5.7116 & 42.88 & 387.66 & 64.12 & 0.12 & 0.08 \\
SSA17    &   
256.70090 & 43.97053 & 5.7275 & 42.95 & 290.28 & 20.51 & 0.33 & 0.02 \\
SSA17    &   
256.73926 & 43.77398 & 5.6626 & 42.86 & 299.08 & 42.69 & 0.39 & 0.04 \\
SSA17    &   
256.76758 & 43.96976 & 5.7012 & 42.74 & 255.60 & 33.75 & 0.11 & 0.06 \\
SSA17    &   
256.78308 & 43.92512 & 5.7010 & 42.67 & 336.17 & 77.82 & 0.25 & 0.08 \\
NEP      &   
259.37943 & 66.38524 & 5.7009 & 43.53 & 205.13 & 8.95 & 0.26 & 0.01 \\
NEP      &   
261.13239 & 66.46720 & 5.6637 & 43.86 & 201.86 & 21.97 & 0.25 & 0.04 \\
NEP      &   
261.20316 & 65.01511 & 5.7403 & 43.48 & 249.22 & 15.50 & 0.29 & 0.02 \\
NEP      &   
261.61444 & 66.34655 & 5.6528 & 44.03 & 271.17 & 11.50 & 0.32 & 0.01 \\
NEP      &   
262.36450 & 68.03432 & 5.7407 & 43.65 & 333.23 & 25.22 & 0.36 & 0.02 \\
NEP      &   
263.36462 & 68.19836 & 5.6957 & 43.48 & 354.00 & 18.06 & 0.33 & 0.02 \\
NEP      &   
263.44412 & 66.33298 & 5.7183 & 43.30 & 294.91 & 42.52 & 0.42 & 0.04 \\
NEP      &   
265.63666 & 68.49759 & 5.7182 & 43.46 & 264.59 & 32.90 & 0.17 & 0.05 \\
NEP      &   
266.47528 & 67.92089 & 5.7040 & 43.39 & 270.96 & 20.63 & 0.26 & 0.03 \\
NEP      &   
266.61945 & 68.01823 & 5.6766 & 43.60 & 250.38 & 17.25 & 0.27 & 0.02 \\
NEP      &   
267.68088 & 66.89002 & 5.7230 & 43.57 & 478.16 & 30.16 & 0.24 & 0.03 \\
NEP      &   
267.78796 & 66.72775 & 5.6963 & 43.49 & 228.42 & 14.40 & 0.22 & 0.02 \\
NEP      &   
267.89209 & 66.62925 & 5.6919 & 43.79 & 505.31 & 43.65 & 0.23 & 0.04 \\
NEP      &   
268.73529 & 64.31248 & 5.7280 & 43.46 & 261.79 & 25.72 & 0.26 & 0.03 \\
NEP      &   
268.74176 & 64.24213 & 5.6950 & 43.45 & 439.56 & 36.47 & 0.18 & 0.04 \\
NEP      &   
270.15982 & 68.81596 & 5.6950 & 43.51 & 369.43 & 77.10 & 0.37 & 0.07 \\
NEP      &   
270.65686 & 67.81599 & 5.7192 & 43.49 & 477.36 & 26.28 & 0.31 & 0.02 \\
NEP      &   
270.96326 & 68.07949 & 5.7250 & 43.55 & 143.85 & 6.06 & 0.09 & 0.02 \\
NEP      &   
271.01611 & 67.01797 & 5.7134 & 43.49 & 307.07 & 27.84 & 0.34 & 0.03 \\
NEP      &   
271.33987 & 67.92043 & 5.7210 & 43.85 & 424.59 & 10.94 & 0.07 & 0.01 \\
NEP      &   
271.72308 & 64.85684 & 5.7061 & 43.31 & 228.82 & 19.48 & 0.32 & 0.02 \\
NEP      &   
271.95343 & 64.92245 & 5.7225 & 43.56 & 239.24 & 17.53 & 0.38 & 0.02 \\
NEP      &   
272.35791 & 64.83362 & 5.7017 & 43.29 & 540.57 & 23.53 & 0.50 & 0.02 \\
NEP      &   
272.41608 & 66.77439 & 5.7448 & 43.48 & 308.91 & 24.75 & 0.25 & 0.03 \\
NEP      &   
272.58472 & 66.69049 & 5.7350 & 43.70 & 408.68 & 11.76 & 0.16 & 0.01 \\
NEP      &   
273.02316 & 68.85086 & 5.7222 & 43.66 & 244.17 & 12.41 & 0.35 & 0.01 \\
NEP      &   
273.24777 & 69.00883 & 5.6995 & 43.25 & 266.41 & 21.12 & 0.12 & 0.04 \\
NEP      &   
274.18512 & 66.97318 & 5.7570 & 43.78 & 309.06 & 8.89 & 0.16 & 0.01 \\
NEP      &   
274.55298 & 67.84969 & 5.7225 & 43.51 & 329.60 & 17.77 & 0.22 & 0.02 \\
NEP      &   
274.67450 & 64.77733 & 5.7218 & 43.44 & 232.47 & 17.02 & 0.35 & 0.02 \\
NEP      &   
274.79449 & 66.03417 & 5.7252 & 43.32 & 262.44 & 18.34 & 0.01 & 0.04 \\
NEP      &   
275.87070 & 64.56598 & 5.7356 & 43.46 & 282.88 & 18.30 & 0.28 & 0.02 \\
NEP      &   
276.02689 & 64.44939 & 5.7170 & 43.42 & 326.69 & 31.89 & 0.17 & 0.04 \\
NEP      &   
276.66409 & 67.53183 & 5.7120 & 43.50 & 233.28 & 11.32 & 0.13 & 0.02 \\
SSA22    &   
333.67297 & -0.26941 & 5.7119 & 43.55 & 185.01 & 4.87 & 0.09 & 0.01 \\
SSA22    &   
333.80704 & -0.49459 & 5.6955 & 43.51 & 262.84 & 17.19 & 0.27 & 0.02 \\
SSA22    &   
334.22006 & 0.27775 & 5.6620 & 42.82 & 328.94 & 67.59 & 0.39 & 0.06 \\
SSA22    &   
334.22888 & 0.09408 & 5.6849 & 43.15 & 364.20 & 18.39 & 0.11 & 0.02 \\
SSA22    &   
334.23529 & 0.24630 & 5.6621 & 42.98 & 357.54 & 75.48 & 0.29 & 0.07 \\
SSA22    &   
334.24414 & 0.14338 & 5.7378 & 42.76 & 128.70 & 30.67 & 0.42 & 0.04 \\
SSA22    &   
334.24490 & 0.31372 & 5.6388 & 42.93 & 425.33 & 81.10 & 0.07 & 0.10 \\
SSA22    &   
334.27319 & 0.21692 & 5.6703 & 42.97 & 327.40 & 33.94 & 0.29 & 0.04 \\
SSA22    &   
334.27838 & 0.20631 & 5.6510 & 42.88 & 276.59 & 65.45 & 0.21 & 0.09 \\
SSA22    &   
334.28033 & 0.46247 & 5.7256 & 42.91 & 304.47 & 58.97 & 0.40 & 0.06 \\
SSA22    &   
334.29242 & 0.21128 & 5.6260 & 43.27 & 429.34 & 163.08 & 0.39 & 0.13 \\
SSA22    &   
334.31812 & 0.22382 & 5.6473 & 43.05 & 585.18 & 75.49 & 0.34 & 0.05 \\
SSA22    &   
334.33691 & 0.33536 & 5.6706 & 43.06 & 167.61 & 26.74 & 0.27 & 0.05 \\
SSA22    &   
334.33722 & 0.29383 & 5.6672 & 43.15 & 460.95 & 36.95 & 0.18 & 0.04 \\
SSA22    &   
334.35437 & 0.13128 & 5.7529 & 42.93 & 277.19 & 18.98 & 0.20 & 0.03 \\
SSA22    &   
334.36969 & 0.32172 & 5.6464 & 43.27 & 259.57 & 25.43 & 0.39 & 0.03 \\
SSA22    &   
334.38171 & 0.16032 & 5.6864 & 42.86 & 237.16 & 32.08 & 0.17 & 0.06 \\
SSA22    &   
334.38794 & 0.37132 & 5.6547 & 43.09 & 204.10 & 18.29 & 0.24 & 0.03 \\
SSA22    &   
334.38995 & 0.70059 & 5.6190 & 43.20 & 310.84 & 47.76 & 0.30 & 0.05 \\
SSA22    &   
334.41321 & 0.42923 & 5.7081 & 42.95 & 382.31 & 56.78 & 0.03 & 0.08 \\
SSA22    &   
334.42044 & 0.40416 & 5.6350 & 43.05 & 318.90 & 31.73 & 0.21 & 0.04 \\
SSA22    &   
334.43582 & 0.67348 & 5.6860 & 42.54 & 333.26 & 50.19 & 0.06 & 0.08 \\
SSA22    &   
334.46979 & 0.60430 & 5.7330 & 42.84 & 253.51 & 21.37 & 0.12 & 0.04 \\
SSA22    &   
334.47122 & 1.11505 & 5.7216 & 43.28 & 200.83 & 16.17 & 0.17 & 0.04 \\
SSA22    &   
334.48886 & 0.68858 & 5.6870 & 42.47 & 215.00 & 49.41 & 0.19 & 0.09 \\
SSA22    &   
334.50912 & 0.24213 & 5.6748 & 43.08 & 235.74 & 21.64 & 0.28 & 0.03 \\
SSA22    &   
334.51050 & 0.50208 & 5.6613 & 42.53 & 375.30 & 75.73 & 0.32 & 0.07 \\
SSA22    &   
334.53766 & 1.12323 & 5.7493 & 43.87 & 165.53 & 6.00 & 0.33 & 0.01 \\
SSA22    &   
334.54785 & 0.08356 & 5.7101 & 43.15 & 183.79 & 12.40 & 0.29 & 0.02 \\
SSA22    &   
334.55170 & 0.72510 & 5.6450 & 42.70 & 293.85 & 54.95 & 0.24 & 0.07 \\
SSA22    &   
334.56860 & 0.61319 & 5.7324 & 42.86 & 205.64 & 21.65 & 0.20 & 0.04 \\
SSA22    &   
334.63617 & 0.47907 & 5.6820 & 43.20 & 250.56 & 10.25 & 0.33 & 0.01 \\
SSA22    &   
334.67990 & 0.74419 & 5.6560 & 43.10 & 175.20 & 14.57 & 0.31 & 0.02 \\
SSA22    &   
334.69003 & 0.56826 & 5.6939 & 42.67 & 256.90 & 27.31 & 0.23 & 0.04 \\
SSA22    &   
334.69302 & 0.53241 & 5.6409 & 42.70 & 213.64 & 24.55 & 0.16 & 0.05 \\
SSA22    &   
334.76755 & 0.42243 & 5.6271 & 42.96 & 225.34 & 79.33 & 0.42 & 0.08 \\
SSA22    &   
334.79910 & -0.51076 & 5.7060 & 43.47 & 205.21 & 10.66 & 0.29 & 0.02 \\
SSA22    &   
334.87347 & -0.93630 & 5.6925 & 43.31 & 250.80 & 22.17 & 0.27 & 0.03 \\
SSA22    &   
334.89648 & -0.78148 & 5.7120 & 43.49 & 206.60 & 12.93 & 0.29 & 0.02 \\
SSA22    &   
335.85464 & -0.69387 & 5.7325 & 43.25 & 154.78 & 33.14 & 0.15 & 0.10 \\
\enddata
\end{deluxetable*}

\startlongtable
\begin{deluxetable*}{ccccccccc}
\tablecaption{$z=6.6$\ Ly$\alpha$ sample  \label{z6tab}}
\tablewidth{350pt}
\tablehead{
\colhead{Field} & \colhead{R.A.} & \colhead{Decl.} &
\colhead{$z$} &
\colhead{$\log {\rm L(Ly}\alpha$)} 
& \colhead{FWHM} & \colhead{error} & \colhead{a$_{asym}$} 
& \colhead{error}\\
\colhead{} & \colhead{} & \colhead{} & \colhead{} &
\colhead{(erg~s$^{-1}$)}  & \multicolumn{2}{c}{$({\rm km\ s}^{-1})$}
}
\startdata
XMM-LSS &   
35.90345 & -3.93370 & 6.5473 & 43.44 & 310.06 & 29.76 & 0.30 & 0.03 \\
XMM-LSS &  
36.09813 & -4.00951 & 6.5634 & 43.61 & 137.78 & 4.80 & 0.27 & 0.01 \\
A370     & 
39.86338 & -1.58983 & 6.4504 & 42.73 & 193.09 & 20.84 & 0.22 & 0.04 \\
A370      &
39.91438 & -1.57567 & 6.5409 & 42.76 & 189.52 & 30.17 & 0.36 & 0.04 \\
A370     & 
39.91479 & -1.58100 & 6.5299 & 42.90 & 160.02 & 16.38 & 0.39 & 0.02 \\
A370      &
39.95666 & -1.52267 & 6.5637 & 42.99 & 221.76 & 21.91 & 0.26 & 0.03 \\
A370      &
39.97811 & -1.55897 & 6.5590 & 43.41 & 227.72 & 18.59 & 0.27 & 0.03 \\
A370      &
40.00753 & -1.68339 & 6.5428 & 42.72 & 182.81 & 39.16 & 0.33 & 0.06 \\
A370      &
40.01770 & -1.38128 & 6.5029 & 42.62 & 112.83 & 40.04 & 0.40 & 0.06 \\
A370      &
40.22926 & -1.72099 & 6.4803 & 42.91 & 300.34 & 49.29 & 0.29 & 0.05 \\
A370      &
40.39424 & -1.61179 & 6.4693 & 42.86 & 233.36 & 37.70 & 0.24 & 0.06 \\
COSMOS    &
150.24178 & 1.80411 & 6.6038 & 43.67 & 268.10 & 9.64 & 0.29 & 0.01 \\
COSMOS    &
150.35336 & 2.52924 & 6.5439 & 43.42 & 277.37 & 19.17 & 0.30 & 0.02 \\
COSMOS    &
150.64742 & 2.20389 & 6.5922 & 43.70 & 248.72 & 10.93 & 0.24 & 0.01 \\
GOODS-N   &
189.35826 & 62.20772 & 6.5593 & 43.45 & 188.57 & 11.81 & 0.22 & 0.02 \\
GOODS-N   &
190.00757 & 62.32957 & 6.5126 & 42.88 & 215.25 & 67.43 & 0.28 & 0.10 \\
GOODS-N   &
190.56474 & 62.29151 & 6.5198 & 42.94 & 241.91 & 62.82 & 0.42 & 0.06 \\
SSA17     &
256.81799 & 43.84443 & 6.5273 & 43.29 & 299.31 & 34.08 & 0.31 & 0.04 \\
NEP       &
259.78857 & 65.38805 & 6.5773 & 43.30 & 386.01 & 49.29 & 0.38 & 0.04 \\
NEP       &
259.91315 & 65.38100 & 6.5752 & 43.45 & 267.95 & 13.11 & 0.21 & 0.02 \\
NEP       &
260.66571 & 66.13013 & 6.5791 & 43.08 & 289.71 & 43.88 & 0.24 & 0.06 \\
NEP       &
260.79163 & 66.06917 & 6.5942 & 43.29 & 233.49 & 21.93 & 0.33 & 0.03 \\
NEP       &
260.80258 & 65.37336 & 6.5815 & 43.48 & 309.97 & 19.40 & 0.22 & 0.02 \\
NEP       &
260.88062 & 65.40775 & 6.5465 & 43.20 & 211.06 & 25.96 & 0.20 & 0.05 \\
NEP       &
261.67026 & 65.83727 & 6.5496 & 43.17 & 228.97 & 27.98 & 0.17 & 0.05 \\
NEP       &
261.70859 & 65.79610 & 6.5617 & 43.40 & 361.73 & 31.45 & 0.24 & 0.03 \\
NEP       &
262.02902 & 66.04409 & 6.5997 & 43.27 & 371.20 & 55.51 & 0.23 & 0.06 \\
NEP       &
262.30841 & 65.59969 & 6.5689 & 43.64 & 312.69 & 20.71 & 0.38 & 0.02 \\
NEP       &
262.44296 & 65.18044 & 6.5670 & 43.78 & 388.33 & 20.24 & 0.27 & 0.02 \\
NEP       &
262.46060 & 65.66861 & 6.5658 & 43.05 & 174.93 & 22.76 & 0.30 & 0.04 \\
NEP       &
263.61490 & 67.59397 & 6.5839 & 43.67 & 213.45 & 8.24 & 0.35 & 0.01 \\
NEP       &
265.22437 & 65.51039 & 6.5989 & 43.67 & 237.56 & 5.36 & 0.27 & 0.01 \\
NEP       &
266.12918 & 68.97475 & 6.5849 & 43.61 & 184.97 & 14.65 & 0.28 & 0.02 \\
NEP       &
266.14337 & 64.72261 & 6.5790 & 43.47 & 286.98 & 43.97 & 0.43 & 0.04 \\
NEP       &
267.42700 & 68.90741 & 6.5500 & 43.47 & 275.83 & 15.01 & 0.23 & 0.02 \\
NEP       &
268.29211 & 65.10958 & 6.5471 & 43.75 & 248.38 & 15.15 & 0.36 & 0.02 \\
NEP       &
269.68964 & 65.94475 & 6.5363 & 43.60 & 334.20 & 25.74 & 0.26 & 0.03 \\
NEP       &
271.92371 & 64.79888 & 6.5501 & 43.37 & 256.95 & 22.12 & 0.31 & 0.03 \\
NEP       &
272.30804 & 67.43481 & 6.5823 & 43.52 & 248.60 & 18.66 & 0.15 & 0.03 \\
NEP       &
272.55881 & 67.26176 & 6.5998 & 43.60 & 342.52 & 25.48 & 0.38 & 0.02 \\
NEP       &
272.66104 & 67.38605 & 6.5784 & 43.59 & 339.27 & 26.18 & 0.39 & 0.02 \\
NEP       &
273.41190 & 68.09296 & 6.5777 & 43.52 & 170.53 & 14.79 & 0.24 & 0.03 \\
NEP       &
273.42078 & 67.51686 & 6.5542 & 43.50 & 350.14 & 26.68 & 0.21 & 0.03 \\
NEP       &
273.73837 & 65.28600 & 6.5795 & 43.94 & 284.99 & 5.88 & 0.35 & 0.01 \\
NEP       &
274.70264 & 64.57021 & 6.5728 & 43.55 & 262.56 & 23.37 & 0.29 & 0.03 \\
NEP       &
274.90005 & 66.12672 & 6.5337 & 43.53 & 376.99 & 30.79 & 0.20 & 0.03 \\
NEP       &
275.76389 & 68.55573 & 6.5504 & 43.52 & 264.44 & 20.89 & 0.32 & 0.02 \\
NEP       &
276.23428 & 67.60670 & 6.5354 & 43.63 & 287.57 & 26.00 & 0.31 & 0.03 \\
SSA22     &
333.68112 & -0.57300 & 6.5582 & 43.33 & 266.45 & 26.02 & 0.36 & 0.03 \\
SSA22     &
334.40994 & 0.15267 & 6.4850 & 42.87 & 183.14 & 35.76 & 0.03 & 0.10 \\
SSA22     &
334.42136 & 0.52625 & 6.5019 & 43.28 & 337.11 & 35.64 & 0.22 & 0.04 \\
SSA22     &
334.42929 & 0.30240 & 6.4702 & 43.20 & 254.62 & 28.63 & 0.10 & 0.06 \\
SSA22     &
334.50800 & 0.81224 & 6.5178 & 42.60 & 172.83 & 15.16 & 0.23 & 0.03 \\
SSA22     &
334.59787 & 0.77540 & 6.4841 & 42.90 & 216.00 & 31.12 & 0.30 & 0.04 \\
SSA22     &
334.61520 & 0.79093 & 6.5051 & 43.05 & 219.20 & 25.58 & 0.15 & 0.05 \\
SSA22     &
334.70264 & 0.73160 & 6.5212 & 42.95 & 186.10 & 33.50 & 0.36 & 0.04 \\
SSA22     &
334.73483 & 0.13537 & 6.5149 & 43.69 & 309.41 & 19.79 & 0.34 & 0.02 \\
SSA22     &
334.74216 & 0.76499 & 6.5556 & 42.81 & 206.76 & 38.15 & 0.26 & 0.06 \\
NEP       &
259.87549 & 66.15894 & 6.5878 & 43.39 & 219.92 & 16.78 & 0.25 & 0.03 \\
NEP       &
260.45947 & 66.05280 & 6.5984 & 43.21 & 368.14 & 86.91 & 0.47 & 0.07 \\
NEP       &
260.75922 & 66.03492 & 6.5937 & 43.00 & 214.71 & 52.50 & 0.36 & 0.06 \\
NEP       &
261.04294 & 65.02969 & 6.5766 & 43.33 & 339.90 & 48.69 & 0.24 & 0.05 \\
NEP       &
261.41815 & 65.31500 & 6.5875 & 43.28 & 191.76 & 189.94 & 0.62 & 0.19 \\
NEP       &
261.46747 & 65.41328 & 6.5930 & 43.17 & 198.52 & 20.60 & 0.14 & 0.05 \\
NEP       &
261.50146 & 65.97598 & 6.5942 & 43.30 & 161.14 & 14.80 & 0.27 & 0.03 \\
NEP       &
261.51166 & 65.96345 & 6.5635 & 43.14 & 323.21 & 82.15 & 0.26 & 0.09 \\
NEP       &
263.04303 & 65.54397 & 6.6018 & 43.37 & 263.27 & 36.18 & 0.28 & 0.04 \\
NEP       &
263.44409 & 66.97511 & 6.6176 & 43.71 & 284.68 & 18.66 & 0.30 & 0.02 \\
NEP       &
263.49301 & 67.46775 & 6.5584 & 43.53 & 400.64 & 73.07 & 0.32 & 0.06 \\
NEP       &
263.56140 & 67.55155 & 6.5502 & 43.32 & 215.68 & 23.62 & 0.31 & 0.03 \\
NEP       &
263.58398 & 67.66872 & 6.5586 & 43.47 & 190.15 & 20.40 & 0.25 & 0.04 \\
NEP       &
267.39105 & 65.91167 & 6.5355 & 43.52 & 261.90 & 11.41 & 0.29 & 0.01 \\
NEP       &
268.06177 & 65.40675 & 6.5856 & 43.45 & 186.88 & 15.72 & 0.25 & 0.03 \\
NEP       &
268.36270 & 68.31772 & 6.5973 & 43.46 & 254.98 & 17.53 & 0.28 & 0.02 \\
NEP       &
270.76889 & 67.26186 & 6.5456 & 43.49 & 230.67 & 28.02 & 0.36 & 0.03 \\
NEP       &
271.02075 & 64.53217 & 6.5989 & 43.39 & 213.44 & 15.63 & 0.24 & 0.03 \\
NEP       &
271.19232 & 64.34397 & 6.5587 & 43.45 & 224.26 & 19.92 & 0.29 & 0.03 \\
NEP       &
273.37930 & 65.41480 & 6.5509 & 43.56 & 320.28 & 20.99 & 0.28 & 0.02 \\
XMM-LSS   &
35.58046 & -3.53547 & 6.5556 & 43.48 & 295.70 & 30.37 & 0.28 & 0.03 \\
NEP       &
267.92221 & 67.10558 & 6.6283 & 43.61 & 196.23 & 26.66 & 0.34 & 0.04 \\
NEP       &
271.40976 & 68.22611 & 6.6165 & 43.46 & 226.94 & 62.48 & 0.36 & 0.07 \\
SSA22     &
334.90829 & -0.07106 & 6.6038 & 43.55 & 317.08 & 18.07 & 0.32 & 0.02 \\
\enddata
\end{deluxetable*}

\begin{table*}[th]
\caption{Spectral Atlas Contents}
\begin{tabular}{ll}\hline\hline
1 & Field \\
2 & Previous Name \\
3 & Current  Name \\
4, 5 & J2000.0 R.A. and Decl. of LAE \\
6 & Redshift of LAE peak \\
7 & Wavelength vector (\AA)\\
8 & Spectrum vector \\
9 & Observed \lya\ luminosity (erg~s$^{-1}$)\\
10 & 2D spectral image centered on \lya\ line \\
11 & Wavelength for 2D spectral image (\AA)\\
12 & Spectrum corresponding to 2D spectral image \\
13 & Velocity centered on \lya\ peak corresponding to 2D spectral image (km~s$^{-1}$) \\
14 & Sky vector corresponding to 2D spectral image \\
15, 16 & FWHM and error (km~s$^{-1}$)\\
17, 18 & Asymmetry paramater ($a_{asym}$) and error\\
19, 20  & Peak amplitude ($A$) and error \\
21, 22 & Width parameter ($d$) and error \\
23 & Double-peaked LAE flag (1=single, 2=double)  \\
24 & Exposure time (hrs)  \\
25 & Date of observation  \\
26--28 &  $15''$ cutout image NB816, $z$, $i$ ($z=5.7$); NB921, $y$, $z$ ($z=6.6$) \\
29--32 &  $2''$ aperture mags NB816, $i$, $z$, $y$ ($z=5.7$); NB921, $i$, $z$, $y$  ($z=6.6$) \\
\hline\hline
\end{tabular}
\label{Atlas}
\end{table*}
\newpage
\clearpage

\bibliography{nsf23.bib}
\end{document}